\documentclass[conference]{IEEEtran}
\IEEEoverridecommandlockouts
\usepackage{cite}
\usepackage{amsmath,amssymb,amsfonts}
\usepackage{algorithmic}
\usepackage{graphicx}
\usepackage{textcomp}
\usepackage{xcolor}

\usepackage[bookmarks=false]{hyperref} % for \url
\IEEEoverridecommandlockouts
\usepackage{tikz}
\usepackage[some,bottom]{background} 
\newcommand\copyrighttext{
 \footnotesize 
 Accepted at 2019 IEEE International Conference on Distributed Computing Systems (ICDCS).
 
 The final authenticated version is available online: DOI 10.1109/ICDCS.2019.00137.
 
 Copyright \textcopyright 2019 IEEE. Personal use of this material is permitted.
  Permission from IEEE must be obtained for all other uses, in any current or future 
  media, including reprinting/republishing this material for advertising or promotional 
  purposes, creating new collective works, for resale or redistribution to servers or 
  lists, or reuse of any copyrighted component of this work in other works. 
 }
\newcommand\copyrightnotice{%
\begin{tikzpicture}[remember picture, overlay]
\node[yshift=10pt]
{\fbox{\parbox{\dimexpr\textwidth-\fboxsep-\fboxrule\relax}{\copyrighttext}}};
\end{tikzpicture}
}
\backgroundsetup{opacity=1, scale=1, angle=0, contents={
\copyrightnotice
}}
\SetBgOpacity{1}
\SetBgScale{1}
\SetBgAngle{0}
\SetBgContents{\copyrightnotice}
\SetBgColor{black}
\usepackage{float}
\usetikzlibrary{patterns}
\usepackage{pgfplots}
\usepackage{pgfplotstable}
\usepgfplotslibrary{
        groupplots,
        statistics,
    }

\usepackage{tabularx,booktabs}
\pgfplotsset{select coords between index/.style 2 args={
    x filter/.code={
        \ifnum\coordindex<#1\fi
        \ifnum\coordindex>#2\fi
    }
}}

\usepackage{pgfplotstable}
\pgfplotstableset{
col sep=semicolon
}

\usetikzlibrary{shapes}
\tikzstyle{block} = [rectangle, draw, 
    text width=5em, text centered, rounded corners, minimum height=2em]
    
\tikzstyle{block2} = [rectangle, draw, 
    text width=6em, text centered, rounded corners, minimum height=2em]

\usepackage[disable]{todonotes}

\def\BibTeX{{\rm B\kern-.05em{\sc i\kern-.025em b}\kern-.08em
    T\kern-.1667em\lower.7ex\hbox{E}\kern-.125emX}}
\begin{document}

\title{Quantitative Impact Evaluation of an Abstraction Layer for Data Stream Processing Systems}

\author{
\IEEEauthorblockN{
Guenter Hesse\IEEEauthorrefmark{1}, 
Christoph Matthies\IEEEauthorrefmark{1}, 
Kelvin Glass\IEEEauthorrefmark{2}, 
Johannes Huegle\IEEEauthorrefmark{1} 
and Matthias Uflacker\IEEEauthorrefmark{1}
} 
\IEEEauthorblockA{
\IEEEauthorrefmark{1}Hasso Plattner Institute\\
University of Potsdam\\
Email: firstname.lastname@hpi.de} 
\IEEEauthorblockA{
\IEEEauthorrefmark{2}Department of Mathematics and Computer Science\\
Freie Universit{\"a}t Berlin\\
Email: kelvin.glass@fu-berlin.de}
}

\maketitle
\BgThispage
%\hspace*{-50pt}
%\copyrightnotice

\begin{abstract}
With the demand to process ever-growing data volumes, a variety of new data stream processing frameworks have been developed.
Moving an implementation from one such system to another, e.g., for performance reasons, requires adapting existing applications to new interfaces.
Apache Beam addresses these high substitution costs by providing an abstraction layer that enables executing programs on any of the supported streaming frameworks.
% ich finde “we propose a benchmark” klingt mist, weil das nicht VLDB-würdig ist. Benchmarks kann jeder schreiben. Besser fände ich “In this paper, we thoroughly evaluate various streaming systems using a single Beam application... oder so”
In this paper, we present a novel benchmark architecture for comparing the performance impact of using Apache Beam on three streaming frameworks: Apache Spark Streaming, Apache Flink, and Apache Apex.
We find significant performance penalties when using Apache Beam for application development in the surveyed systems.
Overall, usage of Apache Beam for the examined streaming applications caused a high variance of query execution times with a slowdown of up to a factor of 58 compared to queries developed without the abstraction layer. 
All developed benchmark artifacts are publicly available to ensure reproducible results.
\end{abstract}

\begin{IEEEkeywords}
Data Stream Processing, Abstraction Layer, Performance Benchmarking
\end{IEEEkeywords}

\section{Introduction}
\label{introduction}

While the world of Big Data analysis presents a multitude of opportunities, it also comes with unique obstacles.
For software developers, who are tasked with building Big Data applications, the need to work with many different frameworks, Application Programming Interfaces (APIs), programming languages, and software development kits (SDKs) can be challenging.
Especially as the long-lasting paradigm "one size fits all" seems to be obsolete, which Stonebraker and \c{C}etintemel~\cite{DBLP:conf/icde/StonebrakerC05} already predicted back in 2005, handling and keeping an overview of the multitude of technologies becomes more difficult.
%Depending on the use case, a given context and the desired goal, developers might choose Apache Flink~\cite{DBLP:journals/debu/CarboneKEMHT} for real-time stream processing, Apache Spark~\cite{DBLP:journals/cacm/ZahariaXWDADMRV16} for batch processing or Google's data stream processing offering running in the cloud that is called Cloud Dataflow~\cite{clouddataflow}.
Furthermore, flexibility is crucial in our rapidly changing world in order to maintain or establish a leader position.
An example for that is a desired change of an IT system for reasons such as variations in pricing policy, changed volume or velocity of data that needs to be processed, or altered performance characteristics of execution engines.
Ideally, this adaption should happen at minimal costs, which is a challenging task if each system uses its own APIs.

The wide variety of available tools in the area of stream processing has spurred the development of an abstraction layer, which allows defining programs independent of the technologies used.
%TODO: ref ORM paper von martin u mir
This layer is the open source project Apache Beam~\cite{beam}, which provides a unified programming model for describing both batch and streaming data-parallel processing pipelines. 
Pipelines are described using a single Software Development Kit (SDK) and can then be executed by a variety of different frameworks, without developers needing detailed knowledge of the employed implementations.
Thus, execution frameworks can be exchanged without the need to adapt code.
As an additional benefit, Apache Beam enables to benchmark multiple systems with a single implementation.  
%This adaptability reduces dependencies on a particular system and allows reacting to changing environments in a flexible manner, which can be helpful if, for example, another DSPS becomes more suitable over time.
%Reasons for this change could be variations in pricing policy, changed volume or velocity of data that needs to be processed, or altered performance characteristics of execution engines.
%This idea is comparable to the ``write once, run anywhere''~\cite{schneider2012nih} concept of Java and the Java Virtual Machine (JVM).
Conceptually, this idea can be compared to object-relational mapping (ORM), where data stored in database tables is encapsulated in objects.
Data can be queried and manipulated just by using these objects instead of writing SQL~\cite{DBLP:conf/hicss/LorenzRHUP17}.
%Conceptually, this idea can be compared to class diagrams of the Unified Modeling Language (UML) modeling standard and model-driven engineering (MDE) techniques~\cite{DBLP:journals/computer/Schmidt06}.
%Instead of implementing a software system in one of a variety of object-oriented programming languages, the structure of a system, its attributes and operations as well as the relationships between actors of the system are described using an abstraction layer.
%Translating these abstract models into program code is then the task of a separate tool.

A question concerning abstraction layers is if their usage has consequences on the performance characteristics of an application.
%Making use of an abstraction certainly has advantages, e.g., with respect to costs for porting programs from one execution frameworks to another.
%Therefore, it is beneficial to quantify the possible performance impact.
%Moreover, comparing different systems with a single implementation of a query or program is very convenient as the development effort for only one implementation is needed.
%The risk of having inaccuracies between measurements due to marginal implementation differences is also eliminated. 
If introduced performance penalties are too high, they might outweigh the gained benefits.
Great performance impact variations between systems can, with regard to performance benchmarking, lead to a distorted result.
%That again could lead to errors of judgement, e.g., with respect to general statements to performance.
Thus, it is crucial to understand and quantify the impact of used abstraction layers. 
The following contributions are presented in this paper:
\begin{itemize}
  \item We give a description of Apache Beam as well as the three data stream processing systems (DSPSs) used for the conducted measurements, which are Apache Flink~\cite{DBLP:journals/debu/CarboneKEMHT}, Apache Spark Streaming~\cite{DBLP:journals/cacm/ZahariaXWDADMRV16}, and Apache Apex~\cite{apex}.
  \item We propose a lightweight benchmark architecture for comparing DSPSs.
  This benchmark covers information on the process as well as data, query, and metric aspects.
  All developed artifacts are available online which ensures transparency and reproducibility.
  \item We present the measurement results with focus on the performance impact of Apache Beam. 
%  We do that by benchmarking three different DSPSs.
  That is done by comparing the query implementations using native system APIs with the implementations using Apache Beam.
\end{itemize}

 %we analyze the performance impact of using an abstraction layer, concretely Apache Beam.
%Therefore, three different DSPSs (DSPSs) are compared.
%As workload, a subset of the queries defined in StreamBench~\cite{DBLP:conf/ucc/LuWXH14} are used. 
%Each query is implemented using the Apache Beam SDK as well as using the native APIs provided by the corresponding system. 

The remainder of this paper is structured as follows:
Section~\ref{technologies} describes the employed technologies.
%In particular, we present Apache Beam as well as the DSPSs used for measurements.
%A comparison of the presented systems concludes Section~\ref{technologies}.
Section~\ref{sec:perfanalysis} illustrates the benchmark environment and the performance results.
Section~\ref{relatedwork}, gives an overview of related work and 
Section~\ref{conclusion} concludes, giving a summary on results and highlighting areas for future work.

\section{Data Stream Processing System Technologies}
\label{technologies}
This section describes the technologies used for the presented measurements.
	
\subsection{Apache Beam} 
\label{beam}

Apache Beam~\cite{beam} describes itself as a unified programming model, which allows defining batch and stream processing applications.
For that, Apache Beam SDKs are provided.
Currently, three SDKs are part of the Apache Beam repository: a Java SDK, a Python SDK, and a Go SDK~\cite{beam_repo}.

Instead of developing an application for a single DSPS, Apache Beam allows writing programs that are compatible with any supported execution engine.
Engine-specific runners translate the Apache Beam code to the target runtime.
Using such an abstraction layer theoretically allows, e.g., for an arbitrary exchange of engines without the need of code adaption.
Central elements of the Apache Beam SDK are:

\begin{itemize}
  \item \textbf{Pipeline} represents the entire application definition, including data input, transformation, and output.
  \item \textbf{PCollection} embodies a distributed data set that can be either bounded or unbounded.
  The latter is used for data stream processing applications.
  \item \textbf{PTranform} is an abstraction for data transformation.
  It receives one or more PCollection objects and applies the a transformation on this data. 
  That leads to an output of zero or more PCollection objects.
  Moreover, read or write operations on external storage systems are realized with PTransform objects.
  Apache Beam provides some core transforms.
  Selected ones are outlined in the following:
  \begin{itemize}
    \item \textbf{ParDo} is an element-by-element processing of data, whereby the processing of a single element can lead to zero or more output elements.
In addition to standard operations like map or flat map, a ParDo also supports aspects such as side inputs and stateful processing.
    \item \textbf{GroupByKey}, as the name already states, processes key-value pairs and collects all values belonging to the same key. 
    It is an aggregation operation that outputs pairs consisting of a key and a collection of values that belong to this key.
    For use with data streams, one must use an aggregation trigger or non-global windowing in order to enable the grouping to be applied to a finite data set.
    %\item \textbf{Combine}
    \item \textbf{Flatten} merges the data of multiple PCollection objects that contain data of the same type into a single PCollection~\cite{beamprogguide,beamrunnerauthguide,DBLP:journals/pvldb/AkidauBCCFLMMPS15}
  \end{itemize}
\end{itemize}

Next to Apache Flink, Apache Spark, and Apache Apex, other frameworks supporting Apache Beam exist~\cite{beamcapa}. 
These are Apache Gearpump~\cite{gearpump}, Apache Hadoop MapReduce~\cite{mapreduce}, Apache Samza~\cite{samza}, Alibaba JStorm~\cite{jstorm}, IBM Streams~\cite{ibmstreams}, and Google Cloud Dataflow~\cite{clouddataflow}.
This group covers both, closed source systems, e.g., Google Cloud Dataflow and IBM Streams, as well as open source systems, e.g., Apache Flink and Apache Spark.
Hence, Apache Beam can be seen as a widely spread project with a high relevance.

Apache Beam itself resulted out of the donation of the Cloud Dataflow SDKs and programming model~\cite{DBLP:journals/pvldb/AkidauBCCFLMMPS15} to the Apache Software Foundation~\cite{cloudbeam}.
Therefore, as mentioned before, Google Cloud Dataflow is one supported system.
%DBLP:conf/bigdataconf/Karau17 -- zur not noch zitieren, aber sagt eig nichts aus

\subsection{Apache Flink} 
\label{flink}

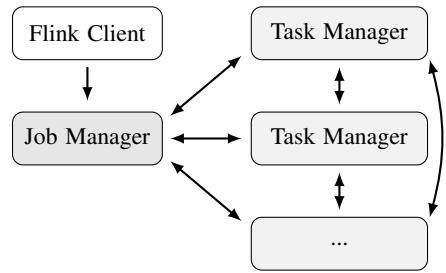
\begin{figure}[!htbp]
\centering
\begin{tikzpicture}
\tikzstyle{every node}=[font=\small]
   \node [block] at (-3, 1.4) {Flink Client};
   \node [block,fill=gray!20] at (-3, 0) {Job Manager};
   \node [block2,fill=gray!10] at (0.35, 1.4) {Task Manager};
   \node [block2,fill=gray!10] at (0.35, 0) {Task Manager};
   \node [block2,fill=gray!10] at (0.35, -1.4) {...};
   
   \draw[>=latex,->, thick] (-3,0.95) -- (-3,0.5);
   \draw[>=latex,<->, thick] (-1.9,0.3) -- (-0.95,1.1);
   \draw[>=latex,<->, thick] (-1.9,0) -- (-0.95,0);
   \draw[>=latex,<->, thick] (-1.9,-0.3) -- (-0.95,-1.1);
   
   \draw[>=latex,<->, thick] (0.35,0.95) -- (0.35,0.45);
   \draw[>=latex,<->, thick] (0.35,-0.95) -- (0.35,-0.45);
   \draw [thick, >=latex, <->] (1.55,1.05) to [bend left=15] (1.55,-1.05);

\end{tikzpicture}
\caption{Architecture of an Apache Flink Runtime (based on\protect\cite{DBLP:journals/debu/CarboneKEMHT,DBLP:conf/icpads/HesseL15})}
\label{flinkarch}
\end{figure}

Apache Flink is an open source system with batch and stream processing capabilities. 
It offers a Java and a Scala API for developing applications.
Additionally, there are multiple libraries on top of Apache Flink that provide, e.g., machine learning or graph processing functionalities~\cite{DBLP:journals/debu/CarboneKEMHT,DBLP:conf/icpads/HesseL15}.
The architecture of an Apache Flink cluster is shown in Figure~\ref{flinkarch}.
\vspace{-2mm}
%wenn mehr infos gewollt, https://ci.apache.org/projects/flink/flink-docs-master/concepts/runtime.html 
%TODO: multiple job managers möglich, siehe doku}

Figure~\ref{flinkarch} shows an Apache Flink Client, a Job Manager, and Task Managers.
When a program is deployed to the system, the client transforms it into a dataflow graph, i.e., a directed acyclic graph (DAG), and sends it to the Job Manager.
The client itself is not part of the program execution and can, after transmitting the dataflow graph, either disconnect from the Job Manager or stay connected in order to receive information about the execution progress.

The Job Manager or master is responsible for scheduling work amongst the Task Manager instances and for keeping track of the execution.
There can be multiple Job Manager instances whereas only one Job Manager can be the leader. 
Others would be standby and could take over in case of failure.  

The Task Manager instances execute the assigned parts of the program.
Technically, a Task Manager is a JVM process.
There must be at least one Task Manager in an Apache Flink installation.
Thereby, they exchange data amongst each other where needed.
Each Task Manager provides at least one task slot in which subtasks are executed in multiple threads.
A task slot can be shared by multiple subtasks as long as they belong to the same application, even if they are part of different tasks.
While one task is executed by one thread, Apache Flink chains multiple operator subtasks into a single task, such as two subsequent map operations.
A benefit of this optimization is, e.g., a reduced overhead for inter-thread communication.
 
Every task slot has a subset of the resources that belong to its corresponding Task Manager.
Particularly, the available memory is split amongst task slots. 
CPU separation does not happen in the current Apache Flink version~\cite{flinkcluster,DBLP:journals/debu/CarboneKEMHT,DBLP:conf/icpads/HesseL15}.

\subsection{Apache Spark Streaming} 
\label{spark}

Apache Spark is another open source system for distributed data processing.
It offers, next to batch processing functionalities, stream processing features as part of its library Apache Spark Streaming.
However, stream processing is implemented using micro-batches, i.e., it is not a tuple-by-tuple processing as in Apache Flink. 
Apache Spark Streaming applications can be written in Java, Scala, or Python.
Besides Apache Spark Streaming, there are other libraries built on top of Apache Spark, e.g., similar to Apache Flink's ecosystem, a library for machine learning as well as for graph processing~\cite{sparkstreamingprogguide,DBLP:journals/cacm/ZahariaXWDADMRV16}.
 
The architecture of an Apache Spark installation is shown in Figure~\ref{spark}.
An application is executed in the form of multiple independent processes distributed across a cluster.
The SparkContext coordinates these processes.
This coordinator is an object in the \textit{main()} function of the application, which is called Driver Program.
Moreover, the SparkContext connects to a Cluster Manager that takes care of resource allocation. 
\vspace{-2mm}
\begin{figure}[!htbp]
%\centering
\begin{tikzpicture}
\tikzstyle{every node}=[font=\small]
   \node[block,
   fill=gray!10,
   text width=6.2em,
   text height=3em] at 
   (-6,0) {};
   \node at (-6,0.3) {Driver Program};
   \node [block,
   fill=gray!30,
   text width=5.75em,
   minimum height=0em,
   text height=0.7em,
   align=center] at 
   (-6, -0.24) 
   {SparkContext};
   \node [block,fill=gray!20] at (-3, 0) {Cluster Manager};
   \node [block,fill=gray!10] at (0.1, 1.4) {Worker Node};
   \node [block,fill=gray!10] at (0.1, 0) {Worker Node};
   \node [block,fill=gray!10] at (0.1, -1.4) {...};
   
   \draw[>=latex,<->, thick] (-2,0.3) -- (-0.9,1);
   \draw[>=latex,<->, thick] (-2,0) -- (-0.9,0);
   \draw[>=latex,<->, thick] (-2,-0.3) -- (-0.9,-1);
   
   \draw[>=latex,<->, thick] (0.1,0.95) -- (0.1,0.45);
   \draw[>=latex,<->, thick] (0.1,-0.95) -- (0.1,-0.45);
   \draw [thick, >=latex, <->] (1.1,0.95) to [bend left=14] (1.1,-0.95);
   
   \draw [thick, >=latex, <->] (-0.9,1.4) to [bend right=15] (-4.8,0);
   \draw [thick, >=latex, <->] (-0.9,-0.4) to [bend left=20] (-4.8,-0.4);
   \draw [thick, >=latex, <->] (-0.9,-1.4) to [bend left=15] (-4.8,-0.65);
   \draw[>=latex,<->, thick] (-4.8,-0.25) -- (-4,0);
\end{tikzpicture}
\caption{Architecture of Apache Spark in Cluster Mode (based on~\protect\cite{spark2,DBLP:conf/icpads/HesseL15})}
\label{spark}
\end{figure}
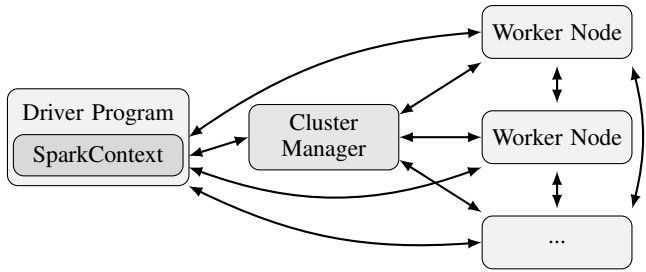

Currently, there are four Cluster Managers supported by Apache Spark - Spark Standalone, Apache Mesos~\cite{hindman2011mesos}, Apache Hadoop YARN (Yet Another Resource Negotiator)~\cite{DBLP:conf/cloud/VavilapalliMDAKEGLSSSCORRB13}, and Kubernetes~\cite{DBLP:conf/cloud/Brewer15}. 
As soon as a connection is established, the SparkContext acquires so-called executors on the Worker Node instances.
Each executor is a process belonging to exactly one application, which stores data and performs computations.
So different applications running on the same Apache Spark Cluster are executed in different JVMs, which is different to, e.g., the  execution concept in the previously illustrated Apache Flink.
Thus, data cannot be exchanged between different Apache Spark application without making use of an external storage system.

Once executors are acquired, the SparkContext transmits the program in the form of a JAR or Python files to them.
Afterwards, it sends tasks to the executor processes. 
One process can run multiple tasks in several threads~\cite{spark2,DBLP:conf/hoti/LuWISP14}. 

A central data structure that is used in Apache Spark is the Resilient Distributed Dataset (RDD).
An RDD can be viewed as a distributed memory abstraction.
To be more concrete, it is a partitioned and read-only collection of records.
Apache Spark Streaming leverages a processing model called discretized streams (D-Streams).
Such a D-Stream is a sequence of RDDs.
An incoming data stream is divided into batches stored in RDDs.
Data transformations are then performed on these RDDs, which again output a D-Stream~\cite{DBLP:conf/nsdi/ZahariaCDDMMFSS12,DBLP:conf/sosp/ZahariaDLHSS13}.
 
\subsection{Apache Apex} 
\label{apex}

Apache Apex is based on Apache Hadoop~\cite{hadoop} with its components Apache Hadoop YARN and Hadoop Distributed File System (HDFS)~\cite{hdfs}.
Similar to Apache Flink, it offers batch as well as stream processing functionalities.
Moreover, stream processing is also implemented in a way of processing data in a tuple-by-tuple fashion~\cite{DBLP:conf/tpctc/Bhandarkar16,dunning2016streaming}.

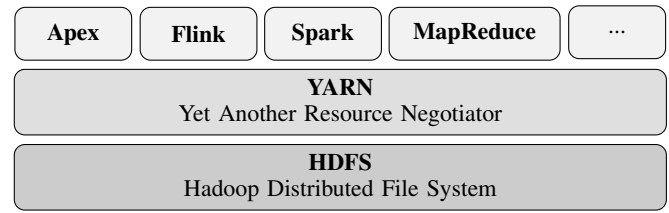
\begin{figure}[!htbp]
\centering
\begin{tikzpicture}
\tikzstyle{every node}=[font=\small]
   
   \node [block,fill=gray!10,text width=3.75em,align=center] at (-3.55, 1.81) {\textbf{Apex}};
   \node [block,fill=gray!10,text width=3.75em,align=center] at (-1.885, 1.81) {\textbf{Flink}};
   \node [block,fill=gray!10,text width=3.75em,align=center] at (-0.23, 1.81) {\textbf{Spark}};
   \node [block,fill=gray!10,text width=5.8em,align=center] at (1.785, 1.81) {\textbf{MapReduce}};
   \node [block,fill=gray!10,text width=3em,align=center] at (3.68, 1.83) {...};
   \node [block,fill=gray!25,text width=24em,align=center] at (0, 0.9) {\textbf{YARN}\\ Yet Another Resource Negotiator};
   \node [block,fill=gray!40,text width=24em,align=center] at (0, -0.1) {\textbf{HDFS}\\ Hadoop Distributed File System};
   
\end{tikzpicture}
\caption{Architecture of Apache Hadoop 2 (based on~\protect\cite{DBLP:conf/sigmod/SahaSSVMC15})}
\label{hadoop}
\vspace{-2mm}
\end{figure}

Apex Malhar, built on top of Apex Core, is a library containing different input/output operators and compute operators.
The former group includes, e.g., connectors to Apache Kafka~\cite{kreps2011kafka} and other messaging systems~\cite{apex}.
The high-level architecture of Apache Hadoop 2, which is the version that is used for the presented measurements, is depicted in Figure~\ref{hadoop}.
HDFS at the bottom is a distributed file system as its name already states and serves as the storage layer.
Apache Hadoop YARN acts as a resource manager on top of HDFS.
On top of YARN, there are multiple data processing frameworks available from various areas such as batch or stream processing.
Two of those are Hadoop MapReduce and Apache Apex. 
The latter is the stream processing system used for measurements described in this paper~\cite{DBLP:conf/sigmod/SahaSSVMC15}.
\vspace{-2mm}

The two other DSPSs used for those measurements, i.e., Apache Spark (Streaming) and Apache Flink, can also run on Apache Hadoop YARN as one of multiple deployment options~\cite{flinkonyarn,sparkonyarn}. 
However, for the presented measurements we use the standalone cluster deployment option that is available in both systems, Apache Flink and Apache Spark.

\begin{figure}[!htbp]
\centering
\begin{tikzpicture}
\tikzstyle{every node}=[font=\small]
   \node [block] 
   at (-3, 1.4) 
   {Client};
   
   \node [
   block,
   fill=gray!10,
   text height=2.3em,
   minimum height=0em
   ] 
   at (-3, 0) 
   {};
   
   \node [
   block,
   fill=gray!30,
   font=\scriptsize,
   text width=3.5em
   ] 
   at (-3, 0) 
   {Resource Manager};
   
   \node [block2,
   fill=gray!10,
   text width=13em,
   text height=1.8em,
   align=right] 
   at (1.4, 1.4) 
   {};
   
   \node [block2,
   fill=gray!30,
   text width=5.6em,
   text height=0.5em,
   font=\scriptsize,
   minimum height=0em,
   ] 
   at (0.4, 1.4) 
   {Node Manager};
   
   \node [block2,
   fill=gray!30,
   text width=5em,
   text height=0.5em,
   text height=0.15cm,
   font=\scriptsize,
   minimum height=0em,
   dashed] 
   at (2.6, 1.4) 
   {Container};
   
   \node [block2,
   fill=gray!10,
   text width=13em,
   text height=1.30cm,
   align=right] 
   at (1.4, -0.08) 
   {};
   
   \node [block2,
   fill=gray!30,
   text width=5.6em,
   text height=0.15cm,
   font=\scriptsize,
   minimum height=0em,
   dashed] 
   at (0.4, 0.29) 
   {App Master};

   \node [block2,
   fill=gray!30,
   text width=5.6em,
   text height=0.5em,
   text height=0.15cm,
   font=\scriptsize,
   minimum height=0em,
   ] 
   at (0.4, -0.45) 
   {Node Manager};
   
   \node [block2,
   fill=gray!30,
   text width=5em,
   text height=0.5em,
   text height=0.15cm,
   font=\scriptsize,
   minimum height=0em,
   dashed] 
   at (2.6, -0.15) 
   {Container};
   
   \node [block2,
   fill=gray!10,
   text width=13em] 
   at (1.4, -1.5) 
   {...};
   
   \draw[>=latex,->, thick] (-3,1) -- (-3,0.42); 
   \draw[>=latex,<->, thick] (-2.25,0.3) -- (-0.75,1.20);
   \draw[>=latex,<->, thick] (-2.25,-0.12) -- (-0.77,-0.45);
   \draw[>=latex,<->, dashed] (-2.25,0.10) -- (-0.77,0.29);
   \draw[>=latex,<->, thick] (-2.25,-0.3) -- (-1,-1.2);
   
   \draw[>=latex,<->, dashed] (1.6,1.2) -- (1.4,0.55);
   \draw [>=latex, <->, dashed] 
   (1.55,0.35) 
   to [bend left=15] 
   (2.2,0.15);

\end{tikzpicture}
\caption{Architecture of an Apache Hadoop YARN (based on\protect\cite{yarn,DBLP:conf/cloud/VavilapalliMDAKEGLSSSCORRB13})}
\label{yarnarch}
\vspace{-4mm}
\end{figure}
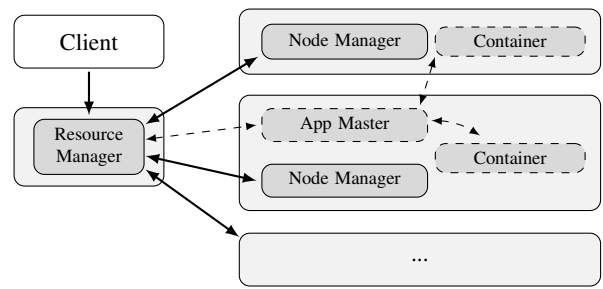

Figure~\ref{yarnarch} illustrates the architecture of an Apache Hadoop YARN deployment.
The depicted installation runs one application.
Its components are marked with dashed lines. 

Two major components are part of Apache Hadoop YARN, a Resource Manager and Node Manager instances.
Both are daemons running on defined nodes. 
A client submits an application to the Resource Manager.
It is responsible for distributing cluster resources amongst applications.
Particularly, the Resource Manager allocates so-called containers on dedicated cluster nodes for applications.
A container is defined as a logical bundle of resources, e.g., a bundle of 4GB RAM and 1 CPU, that is tied to a certain node.
Communication between the Resource Manager and the Node Manager daemons happens via a heartbeat mechanism~\cite{DBLP:conf/cloud/VavilapalliMDAKEGLSSSCORRB13}. 

There is one special container spawned for each program, the Application Master.
It manages application execution with respect to, e.g., resource needs, execution flow, or fault handling. 
The Application Master can be written in any programming language, though many applications make use of higher-level frameworks such as MapReduce or Apache Apex.
Other containers may communicate directly with the Application Master if necessary.
This communication would need to be managed by the application as YARN does not arrange that.
The Application Master implemented in Apache Apex is called Streaming Application Manager (STRAM)~\cite{apexappdevguide,DBLP:conf/cloud/VavilapalliMDAKEGLSSSCORRB13}.

\subsection{Similarities and Differences Between the Presented Systems}
\label{sec:compsystems}

Table~\ref{tab:systemcomp} gives an overview of the presented DSPSs.
Apache Beam is not listed as it is not a DSPS but an SDK for developing stream processing programs.

%\newcolumntype{Y}{>{\centering\arraybackslash}X}
\newcolumntype{b}{X}
\newcolumntype{s}{>{\hsize=.65\hsize}X}
\begin{table}[!htbp]
\begin{tabularx}{\linewidth}{@{}bsss@{}}
\toprule
Criteria & Apache Flink & Apache Spark Streaming & Apache Apex  \\ \midrule
Mainly Written in 
& 
Java, Scala 
& 
Scala, Java, Python 
&
Java
\\
Languages for App Development 
&
Java, Scala, Python
&
Scala, Java, Python
&
Java
\\
Data Processing 
&
Tuple-by-tuple
&
Batch
&
Tuple-by-tuple
\\
Processing Guarantees & 
Exactly-once & 
Exactly-once & 
Exactly-once 
\\
&                                 
&                                            
&              
\\ 
\bottomrule
\end{tabularx}
\caption{Comparison of Apache Flink, Apache Spark Streaming, and Apache Apex (based on~\protect\cite{flinkgithub,apexgithub,sparkgithub,DBLP:conf/icpads/HesseL15})}
\label{tab:systemcomp}
\vspace{-4mm}
\end{table}

All systems are mainly developed in a JVM language, specifically Java or Scala~\cite{flinkgithub,apexgithub,sparkgithub}.
Programs can be written in either Java, Scala, or Python on Apache Flink and Apache Spark Streaming. 
Apache Apex only offers the possibility to write applications in Java.
However, there are plans to also support the development of programs in Scala~\footnote{\url{https://malhar.atlassian.net/browse/APEX-175}}. %TODO: wann footnote, wann quelle?; quellen!!!
With respect to data processing characteristics, Apache Flink and Apache Apex process data in a tuple-by-tuple fashion.
Contrary, Apache Spark Streaming makes use of a batch processing approach.
Furthermore, all three systems guarantee exactly-once processing, i.e., each input tuple is processed exactly once.
This ensures correct results also in recovery scenarios. 

\section{Performance Analysis}
\label{sec:perfanalysis}

This section describes the conducted performance analysis.
First, the general benchmark setup as well as the data used for the benchmark are presented. 
Afterwards, the executed queries are highlighted. 
Lastly, the performance results are discussed.
That particularly includes an analysis of the execution times, standard deviation, and a detailed view of the performance impact of Apache Beam.
Query implementations and other used programs and scripts can be found online~\footnote{\label{githubrepo}\url{http://hpi.de/fileadmin/user_upload/fachgebiete/plattner/publications/papers/gh/StreamBenchOnApacheBeamBenchmark.zip}}.

\subsection{Benchmark Architecture and Process}
\label{sec:setup}
\begin{figure}[!htbp]
\centering
\includegraphics[width=0.47\textwidth]{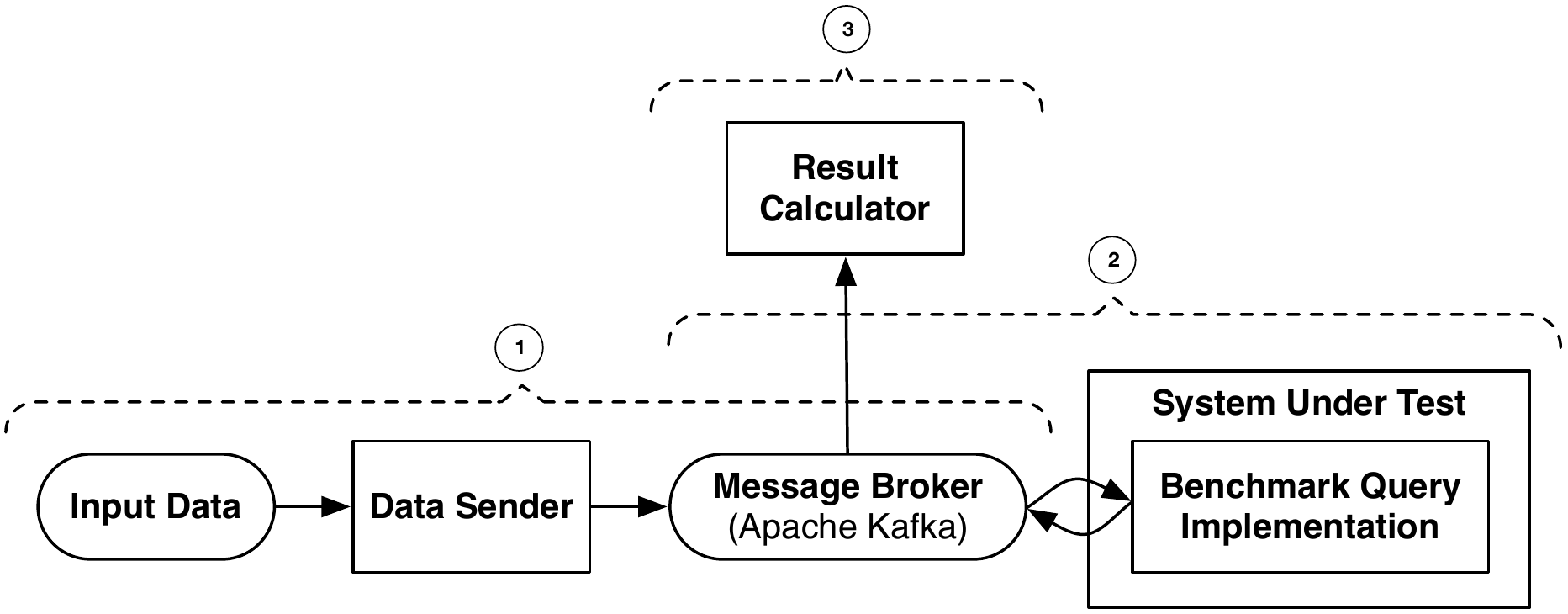}
\caption{Overview About the General Benchmark Architecture and Process Based on Fundamental Modeling Concepts (FMC)}
\label{fig:benchmarkoverview}
\end{figure}

The proposed benchmark setting is depicted in Figure~\ref{fig:benchmarkoverview}.
The benchmark process is divided into three separate and consecutive phases.
The components that are involved in the corresponding part of the process are marked in the figure by dashed curly brackets.
\vspace{-4mm}

On the left-hand side there is the input data which is read by the data sender and forwarded to the message broker, which in particular is Apache Kafka.
The data sender is a program written in Scala with multiple configuration parameters such as the data ingestion rate or the level of Kafka Producer acknowledgments.
The system under test (SUT) on the right-hand side, i.e., the DSPS to be benchmarked, executes the implemented query.
Thereby, it reads from and writes to the message broker.
Moreover, there is a result calculator tool developed in Scala that reads the query output from the message broker and leverages Apache Kafka functionality for the execution time computation.
The three different benchmark process steps marked in Figure~\ref{fig:benchmarkoverview} are described in the following:

\subsubsection{Data Ingestion}
   
  Firstly, data is inserted into an Apache Kafka topic using the data sender.
  Particularly, 1,000,001 records of the AOL Search Query Log~\cite{aoldata} dataset that is also used in~\cite{DBLP:conf/ucc/LuWXH14} are sent.
  The input topic is created with a replication factor of one and one partition in order to ensure the correct order of messages.
  Apache Kafka only guarantees the correct order for entries within one partition~\cite{kafka}.
  The data file consists of records with five tab-separated columns.
  These columns contain a user ID, the query issued by the user, the time at which the query was issued, the search result rank the user clicked on if applicable, and the search result Uniform Resource Locator (URL) the user clicked on if applicable~\cite{aoldata}.
  
\subsubsection{Program Execution}
  
  During the execution phase, each query is run ten times for each execution setup. 
  Meanwhile, there are no other programs executed on the system and each system is restarted at the beginning of this benchmarking step. 
%  Each query implementation uses an Apache Kafka topic that was filled with data before, which acts as query input. 
  The stream processing program computes the output and sends it to an Apache Kafka topic that is also created with a replication factor of one and one partition for the same reasons as mentioned previously. 
  Each query is executed with a parallelism of one and two, ten times each. 
  Moreover, every query is implemented using the APIs provided by the DSPS as well as using Apache Beam.
  So for each query, as we analyzed three different systems, there are twelve different query execution setups as it can be seen in Section~\ref{sec:perfresults}.
  
  The mentioned parallelism is set differently depending on the system and the used APIs.
  Regarding Apache Flink, it is configured using the command line option \emph{-p} or \emph{{-{}-}parallelism} that is offered when submitting an application for specifying parallelism~\cite{flinkcli}. 
  For programs executed on Apache Spark Streaming, the configuration parameter \emph{spark.default.parallelism} is used~\cite{sparkconf}.
  
  Apache Apex does not provide an option for configuring parallelism, so instead the number of VCOREs is set accordingly in the Apache Hadoop YARN configuration~\footnote{\url{http://hadoop.apache.org/docs/stable/hadoop-yarn/hadoop-yarn-common/yarn-default.xml}} as well as within the Apache Apex application as a DAG attribute~\cite{apexoperatorcontext}.
  The approach of using this configuration is also applied for the programs running on Apache Apex that are developed using Apache Beam APIs.
  Details can be found in the mentioned archive that is provided online. 
    
\subsubsection{Result Calculation}
  
   Lastly, the result records are read from Apache Kafka for each query and the time difference between the firstly inserted and the lastly inserted record is computed.
   Apache Kafka is configured to use \textit{LogAppendTime}, i.e., the timestamp when a record is appended to the Apache Kafka log is stored together with the record itself~\cite{kafka}. 
   For execution time calculation, we use these timestamps which allows keeping the measurements application- and system-independent.
   That is a crucial benefit with respect to result correctness as definitions of performance criteria vary among systems.
   Thus, one cannot rely on performance data provided by DSPSs~\cite{DBLP:conf/icde/KarimovRKSHM18}.
   The overhead between having the correct result computed within the SUT and having it appended to Apache Kafka log is identical for every system and hence, results are comparable.

With regard to the hard- and software setup, virtual machines are used for all nodes.
Apache Kafka version 2.11-0.10.1.0 is installed on a three node cluster with 64GB main memory and an Intel(R) Xeon(R) CPU E5-2697 v3 @ 2.60GHz CPU with eight cores each.
The DSPSs are installed on a two node cluster where both nodes act as worker nodes or the equivalent.
These two nodes are identical to the Apache Kafka nodes with regard to both, main memory and CPU.
Ubuntu 14.04 is installed as the operating system on all virtual machines.
Regarding system and framework versions, Apache Apex 3.7.0, Apache Hadoop 2.7.3, Apache Spark 2.3.0, Apache Flink 1.4.0, and Apache Beam 2.3.0 are used.
The configuration files for the different systems can be found in the previously linked archive. 

\subsection{Benchmarked Queries}
\label{sec:queries}

The executed queries are taken from the StreamBench~\cite{DBLP:conf/ucc/LuWXH14} benchmark. 
StreamBench defines seven different queries. 
Four of these are stateless, i.e., it is not required to keep a state for producing the correct answer. 
The remaining three queries are stateful.
The stateless queries used for the benchmark presented in this paper are listed in Table~\ref{querytable}.
Stateful queries are excluded as Apache Beam does not support stateful processing when executed on Apache Spark, see~\cite{beamcapa}.

\begin{table}[!htb]
\begin{tabularx}{\linewidth}{@{}lX}
\toprule
Query      & Description \\ \midrule
Identity   & Read input and output it without performing any data transformation. Can be seen as a baseline query with respect to computational complexity. \\
Sample     & Read input and output only a certain percentage of data that is randomly chosen. The number of output tuples is as big as about 40\% of the number of input tuples. \\
Projection & Read input and output only a certain column of the input record. In the presented measurements, the values of the first column are chosen for being included in the output. \\
Grep       & Read input and output only records that match a certain regex. The search string used for the measurements is "test", which leads to an output of 3,003 records or about 0.3\% of the number of input records. \\ \bottomrule  %out or 1,000,001
\end{tabularx}
\caption{Overview of the Benchmark Queries (based on~\protect\cite{DBLP:conf/ucc/LuWXH14})}
\label{querytable}
\vspace{-8mm}
\end{table}

\subsection{Performance Results}
\label{sec:perfresults}
This section illustrates the performance results regarding execution times and the performance impact of Apache Beam.

\subsubsection{Execution Times}
\label{sec:exectimes}
The following charts visualize the measured average execution times.
On the y-axis, the combinations of system, parallelism, and kind of implementation, i.e., using Apache Beam or system APIs, are listed.
The x-axis shows the times in seconds.
The letter \emph{P} stands for parallelism. 

\begin{figure}[]
\begin{tikzpicture}
\begin{axis}[
    yticklabels from table={results/results_identity.csv}{label},
    y tick label style={rotate=0, font=\small, anchor=east, align=left},
    ytick=data,
    xlabel=Average Execution Time in s,
    axis y line*=left,
    axis x line*=bottom,
    xmin=0,
    width=200pt,
    %xlabel=Number of Run,
    grid=minor,
    %enlargelimits=0.1,
    %legend style={at={(0.5,-0.19)},
    %    anchor=north,
    %    legend columns=-1,
    %   draw=none},
    xbar,
    bar width=7pt,
    bar shift=0pt,
    nodes near coords, nodes near coords align={horizontal}
]
\addplot[
%fill=blue!70,
%select coords between index={1}{1}
] table [
y expr=\coordindex+1, 
x expr=\thisrow{avg}/1000 
] 
{results/results_identity.csv};

\addplot +[
select coords between index={0}{0},
black,
fill=blue!70,
postaction={
  pattern=north east lines
}
] table [
y expr=\coordindex+1, 
x expr=\thisrow{avg}/1000 
] 
{results/results_identity.csv};

\addplot +[
select coords between index={1}{1},
black,
fill=blue!20,
postaction={
  pattern=north east lines
}
] table [
y expr=\coordindex+1, 
x expr=\thisrow{avg}/1000 
] 
{results/results_identity.csv};

\addplot[
fill=blue!70,
select coords between index={2}{2}
] table [
y expr=\coordindex+1, 
x expr=\thisrow{avg}/1000 
] 
{results/results_identity.csv};

\addplot[
fill=blue!20,
select coords between index={3}{3}
] table [
y expr=\coordindex+1, 
x expr=\thisrow{avg}/1000 
] 
{results/results_identity.csv};

\addplot +[
black,
fill=green!70,
postaction={
  pattern=north east lines
},
select coords between index={4}{4}
] table [
y expr=\coordindex+1, 
x expr=\thisrow{avg}/1000 
] 
{results/results_identity.csv};

\addplot +[
select coords between index={5}{5},
black,
fill=green!20,
postaction={
  pattern=north east lines
}
] table [
y expr=\coordindex+1, 
x expr=\thisrow{avg}/1000 
] 
{results/results_identity.csv};

\addplot[
fill=green!70,
select coords between index={6}{6}
] table [
y expr=\coordindex+1, 
x expr=\thisrow{avg}/1000 
] 
{results/results_identity.csv};

\addplot[
fill=green!20,
select coords between index={7}{7}
] table [
y expr=\coordindex+1, 
x expr=\thisrow{avg}/1000 
] 
{results/results_identity.csv};

\addplot +[
black,
fill=orange!70,
postaction={
  pattern=north east lines
},
select coords between index={8}{8}
] table [
y expr=\coordindex+1, 
x expr=\thisrow{avg}/1000 
] 
{results/results_identity.csv};

\addplot +[
select coords between index={9}{9},
black,
fill=orange!20,
postaction={
  pattern=north east lines
}
] table [
y expr=\coordindex+1, 
x expr=\thisrow{avg}/1000 
] 
{results/results_identity.csv};

\addplot[
fill=orange!70,
select coords between index={10}{10}
] table [
y expr=\coordindex+1, 
x expr=\thisrow{avg}/1000 
] 
{results/results_identity.csv};

\addplot[
fill=orange!20,
select coords between index={11}{11}
] table [
y expr=\coordindex+1, 
x expr=\thisrow{avg}/1000 
] 
{results/results_identity.csv};

%\legend{Beam Apex P1,Beam Apex P2}
\end{axis}
\end{tikzpicture}
\caption{Average Execution Times - Identity Query}
\label{fig:identityquery}
%\vspace{-4mm}
\end{figure}
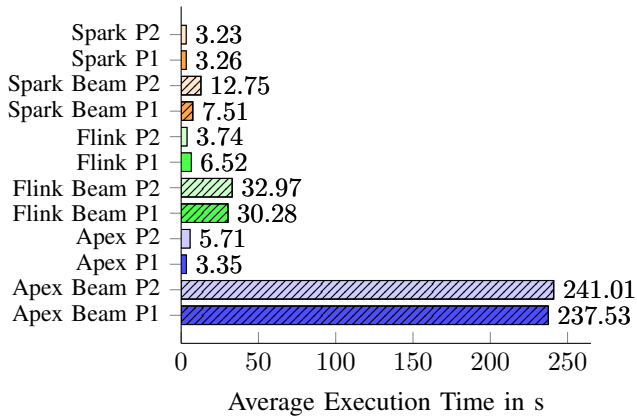

Figure~\ref{fig:identityquery} shows the results for the identity query.
% The average execution times range from 3.23s for Apache Spark Streaming with a parallelism of two to 241.01s for the Apache Beam implementation running on Apache Apex with the same parallelism factor.
It can be seen that the query implementations using Apache Beam are slower compared to the implementations using the APIs provided by the corresponding system in all cases.
That is true for almost all measurements presented in the following.
%It is not a surprising result as using another abstraction layer, e.g., Apache Beam, adds an overhead to applications and reduces the number of optimizations a system can perform.

The overall shortest execution time belongs to queries run on Apache Spark Streaming, closely followed by Apache Apex and Apache Flink, both of which have a noticeable slower average runtime for one kind of parallelism.
That could be due to outliers in the corresponding series of runs.
Details on that can be found in Section~\ref{sec:stddev}.
%However, comparing absolute runtimes between systems is not the goal of this paper as mentioned earlier.

When looking at the runtimes of queries implemented using on Apache Beam, differences are much larger.
%While these queries are again fastest when running on Apache Spark with times of 7.51s and 12.75s for parallelisms of one and two respectively, Apache Flink follows with times between 30s and 33s.
Apache Beam queries running on Apache Apex have by far the highest execution times with around 240s.
So the differences between the execution times of the analyzed systems are significantly higher for the queries implemented using Apache Beam compared to those developed using native system APIs. 
In comparison to these variances, distinctions between parallelism factors are very small.

Nevertheless, the result differences between parallelism factors of the Apache Beam query running on Apache Spark Streaming is noticeable.
The average execution time for the parallelism of two is close to 70\% higher compared to these for the parallelism factor of one.
As the relative standard deviation for these benchmark runs is low as illustrated later on in Figure~\ref{fig:rstddev}, that is not due to outliers.
A reason for this observation could be the introduced overhead with respect to, e.g., data transfer, that comes with splitting up tasks and that may not pay off for simple queries like the identity query.

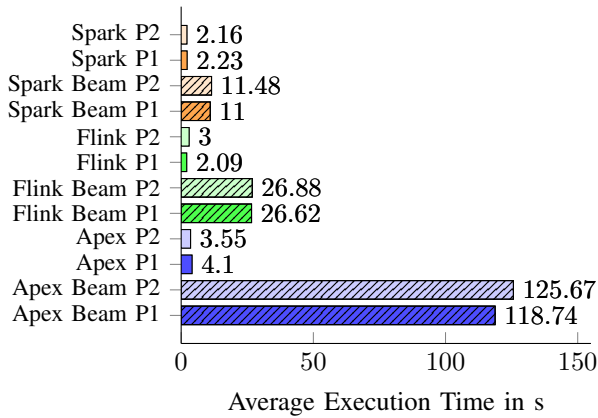
\begin{figure}[]
\begin{tikzpicture}
\begin{axis}[
    yticklabels from table={results/results_sample.csv}{label},
    y tick label style={rotate=0, font=\small, anchor=east, align=left},
    ytick=data,
    xlabel=Average Execution Time in s,
    axis y line*=left,
    axis x line*=bottom,
    xmin=0,
    xmax=155,
    width=200pt,
    %xlabel=Number of Run,
    grid=minor,
    %enlargelimits=0.1,
    %legend style={at={(0.5,-0.19)},
    %    anchor=north,
    %    legend columns=-1,
    %   draw=none},
    xbar,
    bar width=7pt,
    bar shift=0pt,
    nodes near coords, nodes near coords align={horizontal}
]
\addplot[
%fill=blue!70,
%select coords between index={1}{1}
] table [
y expr=\coordindex+1, 
x expr=\thisrow{avg}/1000 
] 
{results/results_sample.csv};

\addplot +[
select coords between index={0}{0},
black,
fill=blue!70,
postaction={
  pattern=north east lines
}
] table [
y expr=\coordindex+1, 
x expr=\thisrow{avg}/1000 
] 
{results/results_sample.csv};

\addplot +[
select coords between index={1}{1},
black,
fill=blue!20,
postaction={
  pattern=north east lines
}
] table [
y expr=\coordindex+1, 
x expr=\thisrow{avg}/1000 
] 
{results/results_sample.csv};

\addplot[
fill=blue!70,
select coords between index={2}{2}
] table [
y expr=\coordindex+1, 
x expr=\thisrow{avg}/1000 
] 
{results/results_sample.csv};

\addplot[
fill=blue!20,
select coords between index={3}{3}
] table [
y expr=\coordindex+1, 
x expr=\thisrow{avg}/1000 
] 
{results/results_sample.csv};

\addplot +[
black,
fill=green!70,
postaction={
  pattern=north east lines
},
select coords between index={4}{4}
] table [
y expr=\coordindex+1, 
x expr=\thisrow{avg}/1000 
] 
{results/results_sample.csv};

\addplot +[
select coords between index={5}{5},
black,
fill=green!20,
postaction={
  pattern=north east lines
}
] table [
y expr=\coordindex+1, 
x expr=\thisrow{avg}/1000 
] 
{results/results_sample.csv};

\addplot[
fill=green!70,
select coords between index={6}{6}
] table [
y expr=\coordindex+1, 
x expr=\thisrow{avg}/1000 
] 
{results/results_sample.csv};

\addplot[
fill=green!20,
select coords between index={7}{7}
] table [
y expr=\coordindex+1, 
x expr=\thisrow{avg}/1000 
] 
{results/results_sample.csv};

\addplot +[
black,
fill=orange!70,
postaction={
  pattern=north east lines
},
select coords between index={8}{8}
] table [
y expr=\coordindex+1, 
x expr=\thisrow{avg}/1000 
] 
{results/results_sample.csv};

\addplot +[
select coords between index={9}{9},
black,
fill=orange!20,
postaction={
  pattern=north east lines
}
] table [
y expr=\coordindex+1, 
x expr=\thisrow{avg}/1000 
] 
{results/results_sample.csv};

\addplot[
fill=orange!70,
select coords between index={10}{10}
] table [
y expr=\coordindex+1, 
x expr=\thisrow{avg}/1000 
] 
{results/results_sample.csv};

\addplot[
fill=orange!20,
select coords between index={11}{11}
] table [
y expr=\coordindex+1, 
x expr=\thisrow{avg}/1000 
] 
{results/results_sample.csv};

%\legend{Beam Apex P1,Beam Apex P2}
\end{axis}
\end{tikzpicture}
\caption{Average Execution Times - Sample Query}
\label{fig:samplequery2}
\vspace{-4mm}
\end{figure}

Figure~\ref{fig:samplequery2} displays the results for the sample query.
Again, it can be seen that implementations using native system APIs outperform these using Apache Beam.
Moreover, results of queries using the system APIs do not differ significantly between the analyzed systems and parallelism factors.
Compared to identity query results, times are slightly lower overall, which could be a result of the lower number of output records as described in Section~\ref{sec:queries}.
%So for, e.g., Apache Spark, average execution times are about 1s lower for the sample query compared to the results for the identity query.
The Apex Beam implementation is an exception as there is a major difference.
To be more concrete, the average execution times for the sample query amount to only about 50\% of the identity query times.

With average execution times of about 2.09s and 3s for the sample query developed using Apache Flink APIs for parallelisms of one and two respectively, these numbers are below the corresponding times for Apache Apex. 
Thus, the performance ranking between systems is identical for the sample query. 
That means, the times for Apache Spark are lowest, followed by these of Apache Flink and Apache Apex for both kinds of implementation.

\begin{figure}[]
\begin{tikzpicture}
\begin{axis}[
    yticklabels from table={results/results_projection.csv}{label},
    y tick label style={rotate=0, font=\small, anchor=east, align=left},
    ytick=data,
    xlabel=Average Execution Time in s,
    axis y line*=left,
    axis x line*=bottom,
    xmin=0,
    width=200pt,
    %xlabel=Number of Run,
    grid=minor,
    %enlargelimits=0.1,
    %legend style={at={(0.5,-0.19)},
    %    anchor=north,
    %    legend columns=-1,
    %   draw=none},
    xbar,
    bar width=7pt,
    bar shift=0pt,
    nodes near coords, nodes near coords align={horizontal}
]

\addplot[
%fill=blue!70,
%select coords between index={1}{1}
] table [
y expr=\coordindex+1, 
x expr=\thisrow{avg}/1000 
] 
{results/results_projection.csv};

\addplot +[
select coords between index={0}{0},
black,
fill=blue!70,
postaction={
  pattern=north east lines
}
] table [
y expr=\coordindex+1, 
x expr=\thisrow{avg}/1000 
] 
{results/results_projection.csv};

\addplot +[
select coords between index={1}{1},
black,
fill=blue!20,
postaction={
  pattern=north east lines
}
] table [
y expr=\coordindex+1, 
x expr=\thisrow{avg}/1000 
] 
{results/results_projection.csv};

\addplot[
fill=blue!70,
select coords between index={2}{2}
] table [
y expr=\coordindex+1, 
x expr=\thisrow{avg}/1000 
] 
{results/results_projection.csv};

\addplot[
fill=blue!20,
select coords between index={3}{3}
] table [
y expr=\coordindex+1, 
x expr=\thisrow{avg}/1000 
] 
{results/results_projection.csv};

\addplot +[
black,
fill=green!70,
postaction={
  pattern=north east lines
},
select coords between index={4}{4}
] table [
y expr=\coordindex+1, 
x expr=\thisrow{avg}/1000 
] 
{results/results_projection.csv};

\addplot +[
select coords between index={5}{5},
black,
fill=green!20,
postaction={
  pattern=north east lines
}
] table [
y expr=\coordindex+1, 
x expr=\thisrow{avg}/1000 
] 
{results/results_projection.csv};

\addplot[
fill=green!70,
select coords between index={6}{6}
] table [
y expr=\coordindex+1, 
x expr=\thisrow{avg}/1000 
] 
{results/results_projection.csv};

\addplot[
fill=green!20,
select coords between index={7}{7}
] table [
y expr=\coordindex+1, 
x expr=\thisrow{avg}/1000 
] 
{results/results_projection.csv};

\addplot +[
black,
fill=orange!70,
postaction={
  pattern=north east lines
},
select coords between index={8}{8}
] table [
y expr=\coordindex+1, 
x expr=\thisrow{avg}/1000 
] 
{results/results_projection.csv};

\addplot +[
select coords between index={9}{9},
black,
fill=orange!20,
postaction={
  pattern=north east lines
}
] table [
y expr=\coordindex+1, 
x expr=\thisrow{avg}/1000 
] 
{results/results_projection.csv};

\addplot[
fill=orange!70,
select coords between index={10}{10}
] table [
y expr=\coordindex+1, 
x expr=\thisrow{avg}/1000 
] 
{results/results_projection.csv};

\addplot[
fill=orange!20,
select coords between index={11}{11}
] table [
y expr=\coordindex+1, 
x expr=\thisrow{avg}/1000 
] 
{results/results_projection.csv};

%\legend{Beam Apex P1,Beam Apex P2}
\end{axis}
\end{tikzpicture}
\caption{Average Execution Times - Projection Query}
\label{fig:projectionquery}
\vspace{-4mm}
\end{figure}
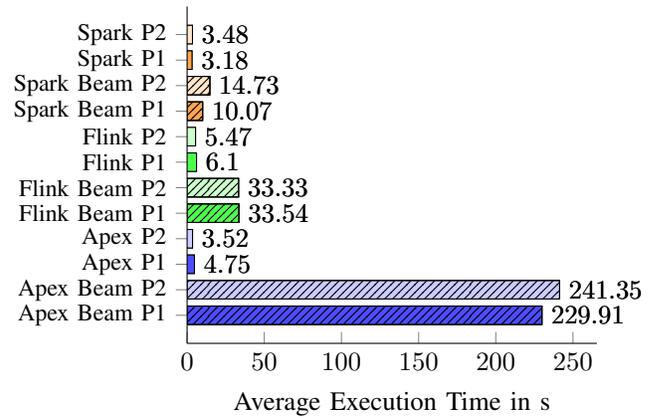

The projection query results are shown in Figure~\ref{fig:projectionquery}.
They are similar to the numbers for the identity query in all aspects.
This closeness leads to the conclusion that splitting a string and accessing one column of the resulting list does not introduce a noticeable overhead.
Regarding the number of output tuples, both queries are identical, though the tuple size for the projection query is smaller as only a subset of columns is sent to the output topic.
However, this reduction in output size does not have a noticeable impact on the times either.

\begin{figure}[]
\begin{tikzpicture}
\begin{axis}[
    yticklabels from table={results/results_grep.csv}{label},
    y tick label style={rotate=0, font=\small, anchor=east, align=left},
    ytick=data,
    xlabel=Average Execution Time in s,
    axis y line*=left,
    axis x line*=bottom,
    xmin=0,
    width=200pt,
    %xlabel=Number of Run,
    grid=minor,
    %enlargelimits=0.1,
    %legend style={at={(0.5,-0.19)},
    %    anchor=north,
    %    legend columns=-1,
    %   draw=none},
    xbar,
    bar width=7pt,
    bar shift=0pt,
    nodes near coords, nodes near coords align={horizontal}
]

\addplot[
%fill=blue!70,
%select coords between index={1}{1}
] table [
y expr=\coordindex+1, 
x expr=\thisrow{avg}/1000 
] 
{results/results_grep.csv};

\addplot +[
select coords between index={0}{0},
black,
fill=blue!70,
postaction={
  pattern=north east lines
}
] table [
y expr=\coordindex+1, 
x expr=\thisrow{avg}/1000 
] 
{results/results_grep.csv};

\addplot +[
select coords between index={1}{1},
black,
fill=blue!20,
postaction={
  pattern=north east lines
}
] table [
y expr=\coordindex+1, 
x expr=\thisrow{avg}/1000 
] 
{results/results_grep.csv};

\addplot[
fill=blue!70,
select coords between index={2}{2}
] table [
y expr=\coordindex+1, 
x expr=\thisrow{avg}/1000 
] 
{results/results_grep.csv};

\addplot[
fill=blue!20,
select coords between index={3}{3}
] table [
y expr=\coordindex+1, 
x expr=\thisrow{avg}/1000 
] 
{results/results_grep.csv};

\addplot +[
black,
fill=green!70,
postaction={
  pattern=north east lines
},
select coords between index={4}{4}
] table [
y expr=\coordindex+1, 
x expr=\thisrow{avg}/1000 
] 
{results/results_grep.csv};

\addplot +[
select coords between index={5}{5},
black,
fill=green!20,
postaction={
  pattern=north east lines
}
] table [
y expr=\coordindex+1, 
x expr=\thisrow{avg}/1000 
] 
{results/results_grep.csv};

\addplot[
fill=green!70,
select coords between index={6}{6}
] table [
y expr=\coordindex+1, 
x expr=\thisrow{avg}/1000 
] 
{results/results_grep.csv};

\addplot[
fill=green!20,
select coords between index={7}{7}
] table [
y expr=\coordindex+1, 
x expr=\thisrow{avg}/1000 
] 
{results/results_grep.csv};

\addplot +[
black,
fill=orange!70,
postaction={
  pattern=north east lines
},
select coords between index={8}{8}
] table [
y expr=\coordindex+1, 
x expr=\thisrow{avg}/1000 
] 
{results/results_grep.csv};

\addplot +[
select coords between index={9}{9},
black,
fill=orange!20,
postaction={
  pattern=north east lines
}
] table [
y expr=\coordindex+1, 
x expr=\thisrow{avg}/1000 
] 
{results/results_grep.csv};

\addplot[
fill=orange!70,
select coords between index={10}{10}
] table [
y expr=\coordindex+1, 
x expr=\thisrow{avg}/1000 
] 
{results/results_grep.csv};

\addplot[
fill=orange!20,
select coords between index={11}{11}
] table [
y expr=\coordindex+1, 
x expr=\thisrow{avg}/1000 
] 
{results/results_grep.csv};

%\legend{Beam Apex P1,Beam Apex P2}
\end{axis}
\end{tikzpicture}
\caption{Average Execution Times - Grep Query}
\label{fig:grepquery}
\vspace{-4mm}
\end{figure}
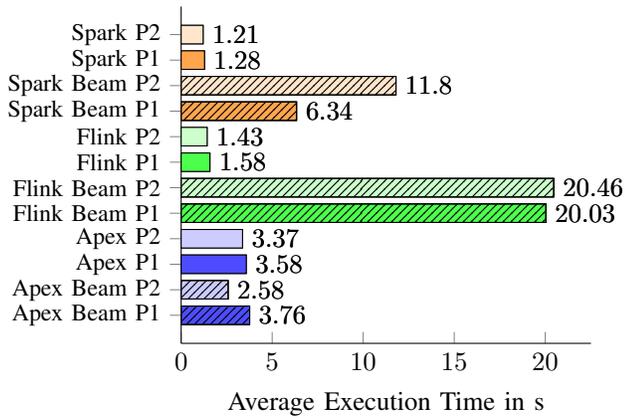

Figure~\ref{fig:grepquery} visualizes the results for the grep query measurements.
These times are overall the lowest ones whereas there are differences between systems and used APIs. 
Especially the implementations using the native APIs offered by Apache Spark Streaming and Apache Flink have relatively low execution times in comparison to the corresponding numbers for the other three queries.
With about 20s, the Apache Beam version for Apache Flink is close to 7s faster than the corresponding sample query result with about 27s.
There is no noticeable difference between parallelism factors.

Contrary to Apache Flink, the average execution times for queries developed using Apache Beam and running on Apache Spark Streaming differ amongst parallelism factors.
Similar to the corresponding times for the identity query depicted in Figure~\ref{fig:identityquery}, the average execution time for a parallelism factor of two is noticeably higher.
In particular, with absolute times of about 11.8s and 6.34s, a parallelism factor of two slows down the average execution time by more than 85\% in comparison to the time measured for a parallelism of one.
Reasons for that can be as described for the identity query results.

A surprising result is the Apex Beam performance for Apache Apex. 
While the times for the native Apache Apex implementation is about on the same level as the corresponding results for all other queries, the ones for the query developed using Apache Beam are drastically lower.
For the projection and the identity query, Apex Beam results are approximately between 230s and 240s.
With about 120s, the sample query performance is already significantly better.
However, with 2.58s and 3.76s, the execution times for the Apex Beam grep query implementation are orders of magnitude lower.

A reason for the relatively low execution times could lie in the number of output records and the resulting smaller effort that is needed for emitting query outcomes.
To be more concrete, the output for the grep query is significantly lower than for the other three queries, though, as described in Section~\ref{sec:queries}, the sample query already outputs fewer tuples than the projection and the identity query.

\subsubsection{Standard Deviation in Measured Execution Times}
\label{sec:stddev}

\begin{figure}[!htb]
\begin{tikzpicture}
  \begin{axis}[
  xbar=0,
  yticklabels from table={results/results_standard_deviation_allgrouped.csv}{label},
  bar width=7pt,
  width=175pt,
  height=300pt,
  ymin=0.3,
  ymax=24.5,
  bar shift=0pt,
  y tick label style={rotate=0, font=\small, anchor=east, align=left},
  ytick=data,
  nodes near coords, 
  nodes near coords align={horizontal},
  every node near coord/.append style={font=\small},
  xlabel=Relative Standard Deviation,
    axis y line*=left,
    axis x line*=bottom,
    xmin=0,
    %xlabel=Number of Run,
    grid=minor
  ]
    \addplot table[x=relstddev, y expr=\coordindex+1] {results/results_standard_deviation_allgrouped.csv};
    %\addplot table[x=rounds, y=QRC] {results/results_standard_deviation.csv};  
    
    \addplot +[
	select coords between index={0}{0},
	black,
	fill=blue!40,
	postaction={
	  pattern=north east lines
	}
	] table [
	y expr=\coordindex+1, 
	x expr=\thisrow{relstddev} 
	] 
	{results/results_standard_deviation_allgrouped.csv};
	
	\addplot +[
	select coords between index={1}{1},
	black,
	fill=blue!40,
	postaction={
	  pattern=north east lines
	}
	] table [
	y expr=\coordindex+1, 
	x expr=\thisrow{relstddev} 
	] 
	{results/results_standard_deviation_allgrouped.csv};
   
   \addplot +[
	select coords between index={2}{2},
	black,
	fill=blue!40,
	postaction={
	  pattern=north east lines
	}
	] table [
	y expr=\coordindex+1, 
	x expr=\thisrow{relstddev} 
	] 
	{results/results_standard_deviation_allgrouped.csv};
	
	\addplot +[
	select coords between index={3}{3},
	black,
	fill=blue!40,
	postaction={
	  pattern=north east lines
	}
	] table [
	y expr=\coordindex+1, 
	x expr=\thisrow{relstddev} 
	] 
	{results/results_standard_deviation_allgrouped.csv};
	
	\addplot +[
	select coords between index={4}{4},
	black,
	fill=blue!40,
	%postaction={
	%  pattern=north east lines
	%}
	] table [
	y expr=\coordindex+1, 
	x expr=\thisrow{relstddev} 
	] 
	{results/results_standard_deviation_allgrouped.csv};
	
	\addplot +[
	select coords between index={5}{5},
	black,
	fill=blue!40,
	%postaction={
	%  pattern=north east lines
	%}
	] table [
	y expr=\coordindex+1, 
	x expr=\thisrow{relstddev} 
	] 
	{results/results_standard_deviation_allgrouped.csv};
   
   \addplot +[
	select coords between index={6}{6},
	black,
	fill=blue!40,
	%postaction={
	%  pattern=north east lines
	%}
	] table [
	y expr=\coordindex+1, 
	x expr=\thisrow{relstddev} 
	] 
	{results/results_standard_deviation_allgrouped.csv};
	
	\addplot +[
	select coords between index={7}{7},
	black,
	fill=blue!40,
	%postaction={
	%  pattern=north east lines
	%}
	] table [
	y expr=\coordindex+1, 
	x expr=\thisrow{relstddev} 
	] 
	{results/results_standard_deviation_allgrouped.csv};

    \addplot +[
	select coords between index={0}{0},
	black,
	fill=blue!40,
	postaction={
	  pattern=north east lines
	}
	] table [
	y expr=\coordindex+1, 
	x expr=\thisrow{relstddev} 
	] 
	{results/results_standard_deviation_allgrouped.csv};
	
	\addplot +[
	select coords between index={1}{1},
	black,
	fill=blue!40,
	postaction={
	  pattern=north east lines
	}
	] table [
	y expr=\coordindex+1, 
	x expr=\thisrow{relstddev} 
	] 
	{results/results_standard_deviation_allgrouped.csv};
   
   \addplot +[
	select coords between index={2}{2},
	black,
	fill=blue!40,
	postaction={
	  pattern=north east lines
	}
	] table [
	y expr=\coordindex+1, 
	x expr=\thisrow{relstddev} 
	] 
	{results/results_standard_deviation_allgrouped.csv};
	
	\addplot +[
	select coords between index={3}{3},
	black,
	fill=blue!40,
	postaction={
	  pattern=north east lines
	}
	] table [
	y expr=\coordindex+1, 
	x expr=\thisrow{relstddev} 
	] 
	{results/results_standard_deviation_allgrouped.csv};
	
	\addplot +[
	select coords between index={4}{4},
	black,
	fill=blue!40,
	%postaction={
	%  pattern=north east lines
	%}
	] table [
	y expr=\coordindex+1, 
	x expr=\thisrow{relstddev} 
	] 
	{results/results_standard_deviation_allgrouped.csv};
	
	\addplot +[
	select coords between index={5}{5},
	black,
	fill=blue!40,
	%postaction={
	%  pattern=north east lines
	%}
	] table [
	y expr=\coordindex+1, 
	x expr=\thisrow{relstddev} 
	] 
	{results/results_standard_deviation_allgrouped.csv};
   
   \addplot +[
	select coords between index={6}{6},
	black,
	fill=blue!40,
	%postaction={
	%  pattern=north east lines
	%}
	] table [
	y expr=\coordindex+1, 
	x expr=\thisrow{relstddev} 
	] 
	{results/results_standard_deviation_allgrouped.csv};
	
	\addplot +[
	select coords between index={7}{7},
	black,
	fill=blue!40,
	%postaction={
	%  pattern=north east lines
	%}
	] table [
	y expr=\coordindex+1, 
	x expr=\thisrow{relstddev} 
	] 
	{results/results_standard_deviation_allgrouped.csv};

    \addplot +[
	select coords between index={8}{8},
	black,
	fill=green!40,
	postaction={
	  pattern=north east lines
	}
	] table [
	y expr=\coordindex+1, 
	x expr=\thisrow{relstddev} 
	] 
	{results/results_standard_deviation_allgrouped.csv};
	
	\addplot +[
	select coords between index={9}{9},
	black,
	fill=green!40,
	postaction={
	  pattern=north east lines
	}
	] table [
	y expr=\coordindex+1, 
	x expr=\thisrow{relstddev} 
	] 
	{results/results_standard_deviation_allgrouped.csv};
   
   \addplot +[
	select coords between index={10}{10},
	black,
	fill=green!40,
	postaction={
	  pattern=north east lines
	}
	] table [
	y expr=\coordindex+1, 
	x expr=\thisrow{relstddev} 
	] 
	{results/results_standard_deviation_allgrouped.csv};
	
	\addplot +[
	select coords between index={11}{11},
	black,
	fill=green!40,
	postaction={
	  pattern=north east lines
	}
	] table [
	y expr=\coordindex+1, 
	x expr=\thisrow{relstddev} 
	] 
	{results/results_standard_deviation_allgrouped.csv};
	
	\addplot +[
	select coords between index={12}{12},
	black,
	fill=green!40,
	%postaction={
	%  pattern=north east lines
	%}
	] table [
	y expr=\coordindex+1, 
	x expr=\thisrow{relstddev} 
	] 
	{results/results_standard_deviation_allgrouped.csv};
	
	\addplot +[
	select coords between index={13}{13},
	black,
	fill=green!40,
	%postaction={
	%  pattern=north east lines
	%}
	] table [
	y expr=\coordindex+1, 
	x expr=\thisrow{relstddev} 
	] 
	{results/results_standard_deviation_allgrouped.csv};
   
   \addplot +[
	select coords between index={14}{14},
	black,
	fill=green!40,
	%postaction={
	%  pattern=north east lines
	%}
	] table [
	y expr=\coordindex+1, 
	x expr=\thisrow{relstddev} 
	] 
	{results/results_standard_deviation_allgrouped.csv};
	
	\addplot +[
	select coords between index={15}{15},
	black,
	fill=green!40,
	%postaction={
	%  pattern=north east lines
	%}
	] table [
	y expr=\coordindex+1, 
	x expr=\thisrow{relstddev} 
	] 
	{results/results_standard_deviation_allgrouped.csv};	
	
	\addplot +[
	select coords between index={16}{16},
	black,
	fill=orange!40,
	postaction={
	  pattern=north east lines
	}
	] table [
	y expr=\coordindex+1, 
	x expr=\thisrow{relstddev} 
	] 
	{results/results_standard_deviation_allgrouped.csv};
	
	\addplot +[
	select coords between index={17}{17},
	black,
	fill=orange!40,
	postaction={
	  pattern=north east lines
	}
	] table [
	y expr=\coordindex+1, 
	x expr=\thisrow{relstddev} 
	] 
	{results/results_standard_deviation_allgrouped.csv};
   
   \addplot +[
	select coords between index={18}{18},
	black,
	fill=orange!40,
	postaction={
	  pattern=north east lines
	}
	] table [
	y expr=\coordindex+1, 
	x expr=\thisrow{relstddev} 
	] 
	{results/results_standard_deviation_allgrouped.csv};
	
	\addplot +[
	select coords between index={19}{19},
	black,
	fill=orange!40,
	postaction={
	  pattern=north east lines
	}
	] table [
	y expr=\coordindex+1, 
	x expr=\thisrow{relstddev} 
	] 
	{results/results_standard_deviation_allgrouped.csv};
	
	\addplot +[
	select coords between index={20}{20},
	black,
	fill=orange!40,
	%postaction={
	%  pattern=north east lines
	%}
	] table [
	y expr=\coordindex+1, 
	x expr=\thisrow{relstddev} 
	] 
	{results/results_standard_deviation_allgrouped.csv};
	
	\addplot +[
	select coords between index={21}{21},
	black,
	fill=orange!40,
	%postaction={
	%  pattern=north east lines
	%}
	] table [
	y expr=\coordindex+1, 
	x expr=\thisrow{relstddev} 
	] 
	{results/results_standard_deviation_allgrouped.csv};
   
   \addplot +[
	select coords between index={22}{22},
	black,
	fill=orange!40,
	%postaction={
	%  pattern=north east lines
	%}
	] table [
	y expr=\coordindex+1, 
	x expr=\thisrow{relstddev} 
	] 
	{results/results_standard_deviation_allgrouped.csv};
	
	\addplot +[
	select coords between index={23}{23},
	black,
	fill=orange!40,
	%postaction={
	%  pattern=north east lines
	%}
	] table [
	y expr=\coordindex+1, 
	x expr=\thisrow{relstddev} 
	] 
	{results/results_standard_deviation_allgrouped.csv};	
	      
  \end{axis}
\end{tikzpicture}
\caption{Relative Standard Deviation for System-Query-SDK Combinations}
\label{fig:rstddev}
\vspace{-4mm}
\end{figure}
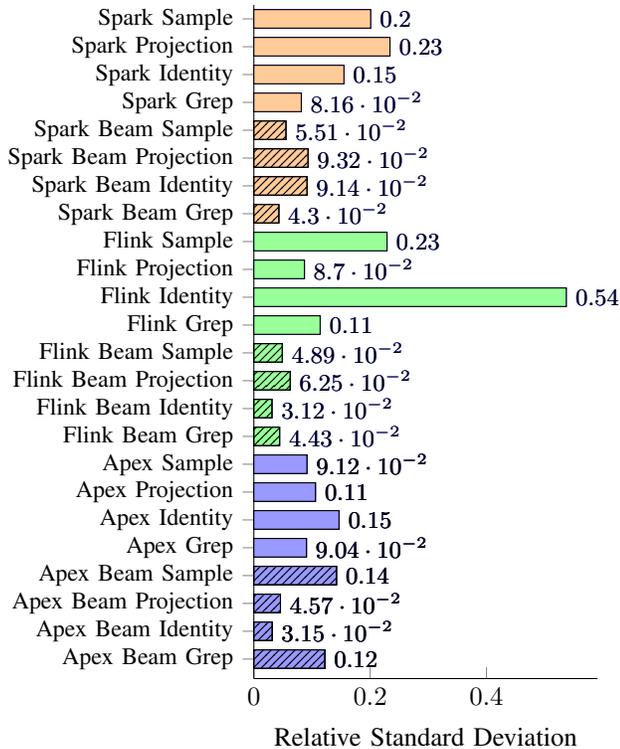

Figure~\ref{fig:rstddev} visualizes the relative standard deviations for the measurements.
These values are calculated for every system-query-SDK combination.
By SDK it is distinguished between using Apache Beam or native system APIs for application development.
Deviations for the two parallelism factors are averaged and condensed in this way.
This is done since separate visualizations for different parallelisms would not reveal any further insights.
Additionally, the reduced number of values simplifies analysis of standard deviations.  

There is one value that is notably higher than others, which belongs to the identity query executed on Apache Flink.
%That is not a surprising result as the absolute standard deviation is already one of the highest while the execution times are relatively low. 
Figure~\ref{fig:identityquery}, which is visualizing the identity query execution times, makes visible that there is a noticeable difference between the numbers for the two parallelism factors for Apache Flink.
Particularly, the execution time for Apache Flink with a parallelism of one is almost 75\% higher than the one for Apache Flink with a parallelism of two.
Although it seems to be plausible at a first glance that higher parallelism leads to better performance, this correlation is absent for other results.

\begin{table}[]
\centering
\begin{tabular}{rrr}
\toprule
\multicolumn{1}{c}{Number of Run} & \multicolumn{1}{c}{Parallelism = 1} & \multicolumn{1}{c}{Parallelism = 2} \\ \midrule
1                                 
& 6.25s
& 4.15s
\\
2 
& 21.56s
& 3.77s
\\
3 
& 3.42s
& 2.71s
\\
4  
& 3.31s
& 5.29s
\\
5  
& 3.73s     
& 3.00s 
\\
6  
& 12.69s 
& 3.93s
\\
7  
& 3.90s
& 2.90s
\\
8  
& 3.96s
& 3.66s
\\
9  
& 3.42s
& 3.57s
\\
10 
& 3.01s
& 4.45s
\\ \bottomrule \\
\end{tabular}
\caption{Execution Times for the Identity Query on Apache Flink}
\label{tab:resflinkidentity}
\vspace{-8mm}
\end{table}
Table~\ref{tab:resflinkidentity} shows the execution times for the benchmark runs of the identity query on Apache Flink, i.e., numbers for the corresponding ten runs with a parallelism of one as well as for the ten runs with a parallelism of two. 
When looking at these measurements it becomes clear that there are two to three outliers that cause this relatively high coefficient of variation.
%The mentioned outliers shown in Table~\ref{tab:resflinkidentity} depict that.
%That lead to the observed relatively high coefficient of variation.
While results for the higher parallelism are relatively homogeneous, there are outliers in the list of execution times for runs with a parallelism of one.
Particularly, seven out of ten execution times range from three to four seconds.
The results for the remaining benchmark runs differ significantly. 
To be more concrete, these runs lasted about 6s, 12.5s, and 21.5s.
The highest execution time, e.g., is more than seven times higher than the lowest one.
These outliers cause the comparatively high relative standard deviation.
Apart from the identity query executed on Apache Flink, there are no further values that stand out in Figure~\ref{fig:rstddev}.

\subsubsection{Performance Impact of Apache Beam}
\label{sec:perfimpactbeam}

The performance impact factors are calculated based on the arithmetic means of the execution times.
These times are measured as defined in Section~\ref{sec:setup}.
The averages are determined as follows: 
\begin{equation*}
	\bar{t}(dsps,query,k,p)=\frac{1}
	{N_{run}}
	\displaystyle\sum_{r=1}^{N_{run}} t(dsps,query,k,p,r)
	,  
\end{equation*}
where $\bar{t}(dsps,query,k,p)$ denotes the average over the execution times for a certain data stream processing system, query, kind of implementation, i.e., using Apache Beam or native system APIs, and a certain parallelism. 
Variable $k$ represents the mentioned kind of implementation and $p$ stands for the used degree of parallelism.
The number of benchmark runs is expressed as $N_{run}$, which is equal to  ten for the context of this paper.
The execution time for a single query run of a certain benchmark scenario is shown as $t(dsps,query,k,p,r)$. 

The slowdown factor calculation makes use of these arithmetic means.
Specifically, it is computed as follows:
\begin{equation*}
	sf(dsps, query) =
	\frac{1}
	{N_{p}} 
	\displaystyle\sum_{p=1}^{N_{p}} 
	\frac{\bar{t}(dsps,query,Beam,p)}
	{\bar{t}(dsps,query,native,p)}
	,  
\end{equation*}
where $sf(dsps,query)$ denotes the slowdown factor for a given data stream processing system and a given query.
$N_{p}$ depicts the number of parallelisms tested, which equals two in the previously discussed benchmark scenario, particularly a parallelism factor of one as well as a parallelism of two.
So in simplified terms, the ratio of average execution times for Apache Beam implementations and these using native system APIs is calculated and again averaged over parallelisms, all for a given query and data stream processing system  combination. 

Concretely, the average execution times for a certain system, query, and parallelism are determined, separately for the Apache Beam version as well as the implementation using native system APIs.
The average execution time belonging to the Apache Beam variant is then divided by the corresponding average for the native query.
That is done for every parallelism. 
The resulting factors for each parallelism are finally averaged by dividing their sum by the number of parallelisms. 

All in all, the result tells how much slower or faster the Apache Beam version for a certain query and data stream processing system performed in the conducted measurements.
That is with regard to execution times as defined in Section~\ref{sec:setup} and independent of parallelism.
A result greater than one marks a slowdown, whereas a result smaller than one means that the Apache Beam implementation was faster than the one using native system APIs.

\begin{figure}[]
\begin{tikzpicture}
\begin{axis}[
    yticklabels from table={results/results_speedup.csv}{system},
    y tick label style={rotate=0, font=\small, anchor=east,align=left},
    ytick=data,
    xlabel=Slowdown Factor $sf_{dsps,query}$,
    axis y line*=left,
    axis x line*=bottom,
    xmin=0,
    width=200pt,
    %xlabel=Number of Run,
    grid=minor,
    %enlargelimits=0.1,
    %legend style={at={(0.5,-0.19)},
    %    anchor=north,
    %    legend columns=-1,
    %   draw=none},
    xbar,
    bar width=7pt,
    bar shift=0pt,
    nodes near coords, nodes near coords align={horizontal}%,
    %every node near coord/.style={font=\scriptsize}
]

\addplot[
%fill=blue!70,
%select coords between index={1}{1}
] table [
y expr=\coordindex+1, 
x expr=\thisrow{factor} 
] 
{results/results_speedup.csv};

\addplot +[
select coords between index={0}{0},
black,
fill=blue!40,
] table [
y expr=\coordindex+1, 
x expr=\thisrow{factor} 
] 
{results/results_speedup.csv};

\addplot +[
select coords between index={1}{1},
black,
fill=blue!40,
] table [
y expr=\coordindex+1, 
x expr=\thisrow{factor} 
] 
{results/results_speedup.csv};

\addplot +[
select coords between index={2}{2},
black,
fill=blue!40,
] table [
y expr=\coordindex+1, 
x expr=\thisrow{factor} 
] 
{results/results_speedup.csv};

\addplot +[
select coords between index={3}{3},
black,
fill=blue!40,
] table [
y expr=\coordindex+1, 
x expr=\thisrow{factor} 
] 
{results/results_speedup.csv};

\addplot +[
select coords between index={4}{4},
black,
fill=green!40,
] table [
y expr=\coordindex+1, 
x expr=\thisrow{factor} 
] 
{results/results_speedup.csv};

\addplot +[
select coords between index={5}{5},
black,
fill=green!40,
] table [
y expr=\coordindex+1, 
x expr=\thisrow{factor} 
] 
{results/results_speedup.csv};

\addplot +[
select coords between index={6}{6},
black,
fill=green!40,
] table [
y expr=\coordindex+1, 
x expr=\thisrow{factor} 
] 
{results/results_speedup.csv};

\addplot +[
select coords between index={7}{7},
black,
fill=green!40,
] table [
y expr=\coordindex+1, 
x expr=\thisrow{factor} 
] 
{results/results_speedup.csv};

\addplot +[
select coords between index={8}{8},
black,
fill=orange!40,
] table [
y expr=\coordindex+1, 
x expr=\thisrow{factor} 
] 
{results/results_speedup.csv};

\addplot +[
select coords between index={9}{9},
black,
fill=orange!40,
] table [
y expr=\coordindex+1, 
x expr=\thisrow{factor} 
] 
{results/results_speedup.csv};

\addplot +[
select coords between index={10}{10},
black,
fill=orange!40,
] table [
y expr=\coordindex+1, 
x expr=\thisrow{factor} 
] 
{results/results_speedup.csv};

\addplot +[
select coords between index={11}{11},
black,
fill=orange!40,
] table [
y expr=\coordindex+1, 
x expr=\thisrow{factor} 
] 
{results/results_speedup.csv};

%\legend{Beam Apex P1,Beam Apex P2}
\end{axis}
\end{tikzpicture}
\caption{Slowdown Factor for the Analyzed Systems and Queries}
\label{fig:speedup}
\vspace{-4mm}
\end{figure}
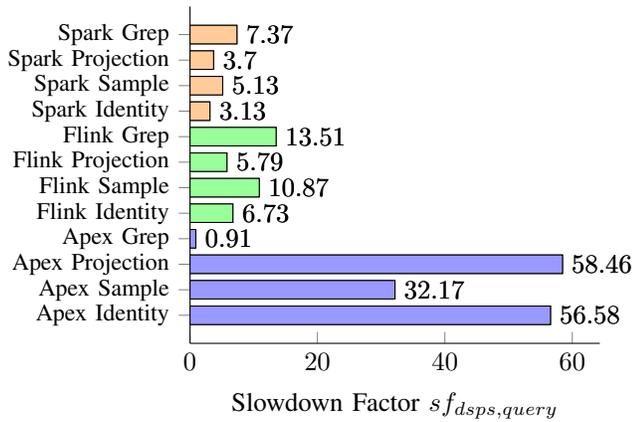

The results for the computed slowdown factors are visualized in Figure~\ref{fig:speedup}. 
It can be seen that Apache Beam implementations are slower for almost all DSPSs and queries in comparison to these developed using native system APIs. 
%That is not a surprising observation as an additional abstraction layer comes with an overhead.

Generally, one can recognize differences between the studied systems and queries.
When looking at Apache Flink and Apache Spark, factors are similar, especially with respect to relative distinctions amongst queries.
Particularly, the performance penalty for the fastest query, namely the grep query, is highest.
Accordingly, it is lowest for the longest-running queries projection and identity for both systems. 
Overall, the performance impact on Apache Flink is slightly higher.

In contrast, the Apache Apex results show a different pattern.
The highest performance impact can be seen, contrary to Apache Flink and Apache Spark, for the longest-running queries projection and identity.
The query with the shortest execution time, the grep query, is overall the only query where the Apache Beam implementation is even faster than the one using native system APIs according to the calculated slowdown factor.
However, this speedup is very low, i.e., it is about as fast as the implementation without Apache Beam.

When looking at the absolute slowdown factors, there are also noticeable differences between Apache Apex and the other two analyzed systems.
Except for the grep query slowdown, all slowdown factors are significantly higher compared to Apache Flink or Apache Spark Streaming.
The slowdown factor for the projection query, e.g., is about 58 and so more than four times higher than the highest slowdown factor for either Apache Flink or Apache Spark Streaming.

Summarizing, the conducted benchmark shows that Apache Beam has a negative performance impact for almost all scenarios.
Averaged over systems, the performance penalty is lowest on Apache Spark, closely followed by Apache Flink. 
Patterns between these two systems and executed queries are similar. 
The performance impact on Apache Apex is much different, meaning the impact is significantly higher in most of the cases and the previously mentioned pattern is vice-versa. 
So the more output or the higher the execution time on Apache Apex, the higher impact of Apache Beam on performance.
The grep query running on Apache Apex is an exception to that as explained before.
Except for this exceptional case, slowdown factors range from about three to almost 60.
Thus, in most of the studied cases, Apache Beam has a significant influence on performance when looking at the calculated slowdown factors. 

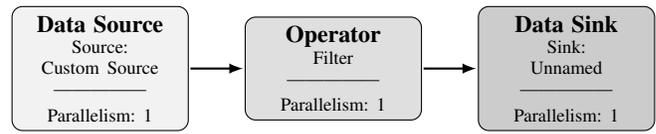
\begin{figure}[]
\centering
\begin{tikzpicture}
\tikzstyle{every node}=[font=\small]
\node [block,fill=gray!10,text width=6em,align=center] at (0, 0) {\textbf{Data Source}\\ \scriptsize Source:\\\scriptsize Custom Source\\---------------\\\scriptsize Parallelism: 1};
\node [block,fill=gray!25,text width=6em,align=center] at (3.1, 0) {\textbf{Operator}\\ \scriptsize Filter \\---------------\\\scriptsize Parallelism: 1};
\node [block,fill=gray!40,text width=6em,align=center] at (6.2, 0) {\textbf{Data Sink}\\ \scriptsize Sink:\\\scriptsize Unnamed\\---------------\\\scriptsize Parallelism: 1};
   \draw[>=latex,->, thick] (1.2,0) -- (1.9,0);
   \draw[>=latex,->, thick] (4.3,0) -- (5,0);
\end{tikzpicture}
\caption{Apache Flink Execution Plan for the Grep Query}
\label{fig:exec_plan_flink_grep}
\vspace{-4mm}
\end{figure}

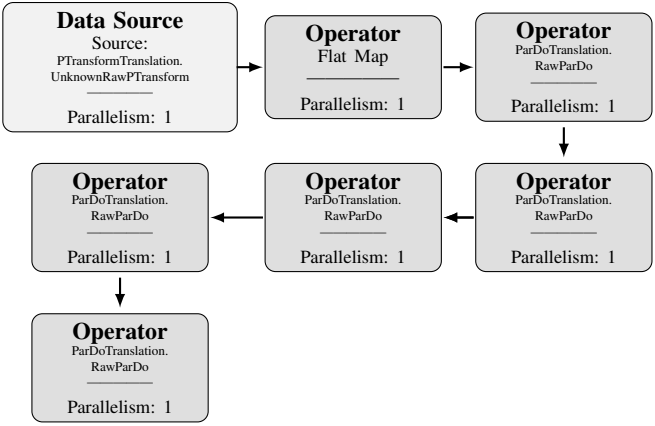
\begin{figure}[]
\centering
\begin{tikzpicture}
\tikzstyle{every node}=[font=\small]
\node [block,fill=gray!10,text width=8.2em,align=center] at (0, 0) {\textbf{Data Source}\\ \scriptsize Source:\\\tiny PTransformTranslation. UnknownRawPTransform\\---------------\\\scriptsize Parallelism: 1};
\node [block,fill=gray!25,text width=6em,align=center] at (3.1, 0) {\textbf{Operator}\\ \scriptsize Flat Map \\---------------\\\scriptsize Parallelism: 1};
\node [block,fill=gray!25,text width=6em,align=center] at (5.9, 0) {\textbf{Operator}\\ \tiny ParDoTranslation. RawParDo\\---------------\\\scriptsize Parallelism: 1};
\node [block,fill=gray!25,text width=6em,align=center] at (5.9, -2) {\textbf{Operator}\\ \tiny ParDoTranslation. RawParDo\\---------------\\\scriptsize Parallelism: 1};
\node [block,fill=gray!25,text width=6em,align=center] at (3.1, -2) {\textbf{Operator}\\ \tiny ParDoTranslation. RawParDo\\---------------\\\scriptsize Parallelism: 1};
\node [block,fill=gray!25,text width=6em,align=center] at (0, -2) {\textbf{Operator}\\ \tiny ParDoTranslation. RawParDo\\---------------\\\scriptsize Parallelism: 1};
\node [block,fill=gray!25,text width=6em,align=center] at (0, -4) {\textbf{Operator}\\ \tiny ParDoTranslation. RawParDo\\---------------\\\scriptsize Parallelism: 1};
   \draw[>=latex,->, thick] (1.55,0) -- (1.9,0);
   \draw[>=latex,->, thick] (4.3,0) -- (4.7,0);
   \draw[>=latex,->, thick] (5.9,-0.8) -- (5.9,-1.2);
   \draw[>=latex,<-, thick] (4.3,-2) -- (4.7,-2);
   \draw[>=latex,<-, thick] (1.2,-2) -- (1.9,-2);
   \draw[>=latex,<-, thick] (4.3,-2) -- (4.7,-2);
   \draw[>=latex,->, thick] (0,-2.8) -- (0,-3.2);
\end{tikzpicture}
\caption{Apache Flink Execution Plan for the Grep Query Implemented Using Apache Beam}
\label{fig:exec_plan_flink_beam_grep}
\vspace{-4mm}
\end{figure}

Figure~\ref{fig:exec_plan_flink_grep} and Figure~\ref{fig:exec_plan_flink_beam_grep} visualize the execution plans for the grep query executed with a parallelism of one on Apache Flink, implemented without and with Apache Beam respectively.
Information on execution plans are retrieved from the Apache Flink system and visualized using the Apache Flink Plan Visualizer~\footnote{\url{https://flink.apache.org/visualizer/}}.
These two plans serve as an example highlighting differences between the execution of  applications developed with native APIs and those using Apache Beam.

The first execution plan depicted in Figure~\ref{fig:exec_plan_flink_grep} contains three elements, a data source, an operator, and a data sink.
Particularly, the source is shown as a custom source, the sink as an unnamed sink, and the operator is a filter, which fits the definition of the grep query as it basically filters data.
Data is forwarded along these three elements.

The second execution plan is presented in Figure~\ref{fig:exec_plan_flink_beam_grep}.
It comprises seven elements in total.
In particular, these elements are a data source followed by six operators.
The data source at the beginning is named PTransformTranslation.UnknownRawPTransform.
PTransformTranslation is a registry of familiar transforms and uniform resource names (URNs)~\cite{beamrunnerauthguide}.
As outlined in Section~\ref{beam}, a PTransform is used, e.g., for reading or writing to an external storage system~\cite{beamprogguide}.

The data source forwards data to the first operator, a flat map, which performs an action on each input value and produces zero or more output values.
Its Apache Beam counterpart is the \textit{read()} method of the \textit{KafkaIO} class, which creates a \textit{Read} PTransform.
%This might be the operator used for applying the logic of the grep query, i.e., only forwarding input tuples that match a certain regex.
The remaining five ParDoTranslation.RawParDo operators follow the flat map.
A ParDoTranslation comprises tools for working with instances of ParDo.
A ParDo is one of the core transforms provided by Apache Beam and described in Section~\ref{beam}.
The first ParDo represents calling \textit{withoutMetadata()} on the \textit{Read} PTransform, which drops the Kafka metadata as it is not needed.
Moreover, the method again returns a PTransform containing a PCollection of key-value pairs.
The downstream operator represents the call of the \textit{create()} method belonging to the class \textit{Values}. 
This operator takes the previously created PCollection of key-value pairs and returns a PCollection containing only the values.
Further downstream, the grep query logic is applied and resulting values are sent to Apache Kafka~\cite{beam_repo,beamprogguide,beamrunnerauthguide,DBLP:journals/pvldb/AkidauBCCFLMMPS15}.
 
When comparing both execution plans it becomes visible that the plan for the query implemented using Apache Beam is significantly larger, i.e., it contains more elements in comparison to its counterpart.
That is due to the more complex management of communication with Apache Kafka and could cause a lower performance. 
%Specifically, the Apache Beam plan is divided into seven parts while the other plan only contains three elements.
Both plans have in common that they start with a data source and that all elements are executed with a parallelism of one, due to the defined degree of parallelism.
Moreover, a dedicated data sink is not identified for the program developed using Apache Beam. 
Thus, the sink must be represented as an operator. 

%Overall, having a larger execution plan for programs developed using Apache Beam is not unexpected as abstraction layers introduce an overhead which could be expressed by that.
%The same holds true for the fuzzier labeling of operators, which is already implied by the name abstraction layer. 

Overall, the performance of Apache Beam applications highly depends on the runner implementations. 
Effort put into this development is likely to vary between systems.
The closeness of the DSPSs' programming model to the underlying concepts of Apache Beam also impacts the application execution.
Further details, e.g., with respect to the concrete impact of the additional operator, could be uncovered through profiling applications. 
However, all measurements are a snapshot in time and results may differ with different versions of Apache Beam, other DSPSs versions, or alternative system configurations.
Moreover, changed workloads characteristics might also influence performance results.

\section{Related Work} 
\label{relatedwork}

For supporting Apache Beam, a system has to implement a so-called runner.
The development of the runner for IBM Streams is described in~\cite{DBLP:journals/pvldb/LiGMDMW18}.
Next to highlighting three implemented optimizations with regard to the IBM Streams runner, performance evaluations between IBM Streams, Apache Flink, and Apache Spark are presented.

Besides abstraction layers such as Apache Beam that allow for developing data stream processing applications using a programming language such as Java, there is the idea of leveraging or extending SQL to be able to do that.

Continuous Query Language (CQL)~\cite{DBLP:journals/vldb/ArasuBW06} is a comprehensive approach for such an SQL extension.
To be more concrete, CQL is a SQL-based language for defining continuous queries over data streams as well as updatable relations.
It is not only a concept but also integrated into the STREAM~\cite{DBLP:journals/debu/ArasuBBDIMNSTVW03} system, a data stream management system developed at Stanford University.
Next to describing the CQL implementation in STREAM, the semantics of CQL are outlined and a comparison to other languages is included in the linked paper.

Apache Calcite~\cite{DBLP:conf/sigmod/BegoliCHML18} is another approach that was developed more recently.
It is a framework that comprises various functionalities, e.g., with respect to query processing, query optimization, and query language support, which is the relevant aspect in this context.
Amongst others, the architecture of Apache Calcite is presented in the mentioned work as well as developed SQL extensions for different areas such as semi-structured data or geospatial queries.
Another depicted example is the extensions for data stream processing queries that are called STREAM extensions.
They are inspired by the mentioned CQL and also explained on their website~\cite{calcitewebside}.
However, not all the presented concepts have been implemented yet~\cite{calcitewebside}.
A few DSPSs already integrate Apache Calcite, Apache Apex and Apache Flink being two of them~\cite{DBLP:conf/sigmod/BegoliCHML18}.

Jain et al.~\cite{DBLP:journals/pvldb/JainMSGWBCCTZ08} discuss the differences of two SQL-based languages for defining streaming queries, particularly Oracle Continuous Query Language~\cite{oraclecql} and StreamBase StreamSQL~\cite{streamsql}.
Moreover, an approach for unifying both languages is proposed.
However, they highlight that for achieving a complete standard, further challenges needs to be tackled.

Besides, there are more SQL extensions for stream processing scenarios developed for certain systems, e.g., streaming SQL for Apache Kafka called KSQL~\cite{ksql}, Continuous Computation Language (CCL) that is the extended SQL used in SAP HANA Smart Data Streaming~\cite{ccl}, or SamzaSQL~\cite{DBLP:conf/ipps/PathirageHPP16} as extended SQL for the DSPS Samza~\cite{Noghabi:2017:SSS:3137765.3137770}.

With respect to benchmarking DSPSs in general, the Linear Road benchmark by Arasu et al.~\cite{DBLP:conf/vldb/ArasuCGMMRST04} is a very well-known work.
It is an application benchmark that provides a benchmarking toolkit.
This toolkit consists of a data generator, a data sender, and a result validator.
The underlying idea of the benchmark is a variable tolling system for a metropolitan area.
This area covers multiple expressways with moving vehicles.
The amount of accumulated tolls depends on various aspects concerning the traffic situation.

The mentioned data sender emits data to the DSPS, which is mostly car position reports.
%This input data contains four different record types, from which position reports are by far the most abundant records. 
%The remaining data consists of three record types that express explicit user requests that always expect an answer from the system. 
%Depending on the overall situation on highways, car position reports may require the SUT to create an output or not.  
Depending on the overall situation on the expressways, car position reports may require the DSPS to create an output or not.  
Next to car position reports, the remaining input data represent an explicit query which always requires an answer. 
Linear Road defines four distinct queries, whereas the query lastly presented in~\cite{DBLP:conf/vldb/ArasuCGMMRST04} was skipped in the two presented implementations due to complexity reasons.
%With regard to the benchmark workload, Linear Road defines four different queries with corresponding output types. 
%For complexity reasons, the implementation of the lastly presented query was even skipped in the two implementations described in~\cite{DBLP:conf/vldb/ArasuCGMMRST04}.
%Besides streaming data, historical data covering ten weeks of tolling history is generated and partly has to be used in order to produce correct answers.

The benchmark result for a system is summarized as a so called L-rating. 
This metric defined by Linear Road expresses how many expressways the system could handle while meeting the defined response time requirements for each query.
A higher number of expressways corresponds to a higher data input rate for the SUT.
When generating data, the amount of expressways can be configured.s
The Linear Road benchmark was applied to the DSPS Aurora~\cite{DBLP:journals/vldb/AbadiCCCCLSTZ03} and a commercial relational database.
Results are presented in the paper.
%As a benchmark result, Linear Road defines one overall metric called L-Rating.
%The L-Rating indicates how many expressways a system can handle without violating the defined maximum response times for each query.
%The number of highways is a configurable parameter for the data generation step that is influencing the amount of input data.

Another related benchmark is the already mentioned StreamBench~\cite{DBLP:conf/ucc/LuWXH14}.
It aims at benchmarking distributed DSPSs.
Regarding its category in the area of performance benchmarks it can be viewed as a microbenchmark, i.e., it measures atomic operations such as a projection rather than more complex applications as in, e.g., Linear Road.
%Thus, when a system's performance for real-world scenarios or applications is to be evaluated, micro benchmark results only have limited validity.
%However, if, e.g., two distinct filter operators are to be compared, micro benchmarks have advantages over application benchmarks due to their simplicity. 
%Measurements contain only the relevant parts without much overhead, which eases interpreting results.

The seven queries defined by StreamBench partly contain a single computational step and partly comprise multiple computational steps. 
Three queries require to keep a state in order to calculate correct results while the remaining four queries are stateless. 
One query uses numerical data as input while the others process textual data.
With respect to the benchmark architecture, StreamBench makes use of a message broker, specifically Apache Kafka, for decoupling data generation and consumption.
That is different to Linear Road but similar to the benchmark architecture proposed in Section~\ref{sec:setup}.
As part of the evaluation section, the DSPSs Apache Storm~\cite{storm} and Apache Spark Streaming are compared by applying StreamBench. 

NEXMark~\cite{nexmark} is a benchmark aiming to benchmark streaming queries in the context of an online auction system.
It seems the benchmark was never finished as the website still states that it is "work in progress"~\cite{nexmark} and there is only a draft version of a paper that was published a couple of years ago~\cite{nexmarkdraft}.
However, in the context of Apache Beam this benchmark was used as inspiration and foundation for a NEXMark-based benchmark suite~\cite{nexmarksuite}.
This suite extends the eight NEXMark queries by five additional ones.
A complete implementation of all queries for all runners is work in progress according to~\cite{nexmarksuite}. 

The work presented in~\cite{DBLP:conf/cluster/MarcuCAP16} compares Apache Flink and Apache Spark.
The conducted measurements include different queries, a grep query being one of them.
One focus area that is analyzed is the scaling behavior with regard to different numbers of nodes in the cluster.
However, studying both systems from a data stream processing point of view is out of scope in the performed measurements.

Lopez et al.~\cite{DBLP:conf/globecom/LopezLD16} compare Apache Storm, Apache Flink, and Apache Spark Streaming in their paper.
Besides describing the architecture of these three systems, the performance is studied in a network traffic analysis scenario.
Additionally, the behavior in case of a node failure is investigated.

%To the best of our knowledge, there is no work published that studies the performance impact of Apache Beam or similar abstraction layers for DSPSs as presented in this paper.

\section{Conclusion and Future Work} 
\label{conclusion}

This paper describes the characteristics of three state-of-the-art DSPSs, particularly Apache Apex, Apache Flink, and Apache Spark Streaming, as well as the abstraction layer Apache Beam for implementing stream processing programs.
%This layer and allows executing programs on different streaming engines.

To study the performance impact of Apache Beam, we propose a lightweight benchmark architecture that uses a recognized workload and a novel approach for measuring execution times using Apache Kafka.
%TODO cite for ``recognized workload''?
%TODO a novel approach for measuring execution times, using... ?
All benchmark implementations are provided in order to ensure reproducibility.

Our benchmark results show that Apache Beam has a noticeable impact on the performance of DSPSs in almost all cases.
Programs developed using Apache Beam suffered from a slowdown of up to a factor of 58 in the worst case.
At the same time, there is one scenario where the query developed using Apache Beam is about as fast as its counterparts using the APIs of the corresponding DSPS.
However, for most scenarios we observed a slowdown of at least a factor three.

The results lead to two major conclusions.
Firstly, using Apache Beam as an abstraction layer for application development comes at a cost in terms of runtime performance.
Secondly, the results of benchmarking different DSPSs using a program developed with Apache Beam are not likely to represent the performance differences that are to be expected from a benchmark with programs developed using native system APIs.
While using Apache Beam certainly provides a greater flexibility to switch underlying DSPSs with relatively low effort, one needs to be aware of the fact that this advantage comes with a negative impact on performance.
This performance penalty varies among systems and applications and is currently unpredictable.

Future work in this area involves studying the reasons for performance differences in greater detail.
Particularly, the applications could be profiled in order to see how much time is spent in which part of the execution plans and thus, to identify possible performance bottlenecks.
Finding potential reasons for the comparatively large performance penalties when employing Apache Apex represents another interesting area for future work.
In the best case, it is possible to identify factors that influence the performance penalty applications suffer from and make them predictable.
Additionally, measurements can be extended with respect to various aspects such as the number of studied systems or query complexity as well as scaling, parallelism, or fault-tolerance behaviors.
Furthermore, upcoming Apache Beam versions and other abstraction layers can be compared against the presented results to supplement the initial overview presented here.

\bibliographystyle{IEEEtran}
\bibliography{icdcs.bib}

\end{document}